\documentclass[review]{elsarticle}

\usepackage{lineno,hyperref}
\modulolinenumbers[5]

\journal{Journal of Computational Physics}

\usepackage{graphicx}
\usepackage{epstopdf, epsfig}
\usepackage[section]{placeins} %Float Barriers
\usepackage{amsmath}
\usepackage{color}
\usepackage{rotating}
\usepackage{lineno,hyperref}
\usepackage{amsmath}
\usepackage{bm}
\usepackage{caption}
\usepackage{subcaption}
\usepackage{float}
\usepackage{tikz, graphicx}
\usepackage{color}
\usetikzlibrary{shapes}
\usepackage{mathtools}
\usepackage{csquotes}
\usepackage{amssymb}
\usepackage{mathrsfs}
\usetikzlibrary{shapes}
\usepackage{mathtools}
\usepackage{xcolor}
\usepackage{amssymb}
\usepackage{bm}
\usepackage{nameref}
\usepackage[toc,page]{appendix}
\usepackage[english]{babel}
\usepackage[T1]{fontenc}
\usepackage{lipsum}
\usepackage{caption}
\usepackage{longtable}

\biboptions{authoryear}

\begin{document}

\begin{frontmatter}

%\title{Elsevier \LaTeX\ template\tnoteref{mytitlenote}}
\title{A fully Eulerian hybrid Immersed Boundary-Phase Field Model for contact line dynamics on complex geometries}

%\author{Ali Yousefi\fnref{myfootnote}}

\author[Stoccolma]{Armin Shahmardi\corref{mycorrespondingauthor}}
\cortext[mycorrespondingauthor]{Corresponding author: arminsh@mech.kth.se}
\author[Okinawa]{Marco Edoardo Rosti}
\author[Stoccolma]{Outi  Tammisola}
\author[Stoccolma,Norway]{Luca Brandt}


\address[Stoccolma]{SeRC (Swedish e-Science Research Centre) and Flow, KTH, \\ Department of Engineering Mechanics, SE-10044 Stockholm, Sweden}
\address[Okinawa]{Complex Fluids and Flows Unit, Okinawa Institute of Science and Technology Graduate University, 1919-1 Tancha, Onna-son, Okinawa 904-0495, Japan}
\address[Norway]{Department of Energy and Process Engineering, Norwegian University of Science and Technology (NTNU), Trondheim, Norway}
%\address[mymainaddress]{Linn\' e FLOW Centre}
%\address[mysecondaryaddress]{Padova}
\begin{abstract}
{We present a fully Eulerian hybrid immersed-boundary/phase-field model to simulate wetting and contact line motion over any arbitrary geometry. The solid wall is described with a volume-penalisation ghost-cell immersed boundary whereas the interface between the two fluids by a diffuse-interface method. The contact line motion on the complex wall is prescribed via slip velocity in the momentum equation and static/dynamic contact angle condition for the order parameter of the Cahn-Hilliard model. This combination requires accurate computations of the normal and tangential gradients of the scalar order parameter and of the components of the velocity. However, the present algorithm requires the computation of averaging weights and other geometrical variables as a preprocessing step. Several validation tests are reported in the manuscript, together with 2D simulations of a droplet spreading over a sinusoidal wall with different contact angles and slip length and a spherical droplet spreading over a sphere, showing that the proposed algorithm is capable to deal with the three-phase contact line motion over any complex wall. The Eulerian feature of the algorithm facilitates the implementation and provides a straight-forward and potentially highly scalable parallelisation. The employed parallelisation of the underlying Navier-Stokes solver can be efficiently used for the multiphase part as well.
The procedure proposed here can be directly employed to impose any types of boundary conditions (Neumann, Dirichlet and mixed) for any field variable evolving over a complex geometry, modelled with an immersed-boundary approach (for instance, modelling deformable biological membranes, red blood cells, solidification, evaporation and boiling, to name a few)}.
\end{abstract}


\end{frontmatter}

\begin{center}
\begin{longtable}{ c  c c }
  Symbols & Definitions & units in (SI) \\
\hline
\it{AP1}&Averaging points for the first interpolation point&$-$\\ 
\it{AP2}&Averaging points for the second interpolation point&$-$\\ 
\it{C}& Concentration&$-$\\ 
\it{$Ca$}&Capillary number&$-$\\  
\it{$Cn$}& Cahn number&$-$\\     
\it{$d_n$}&Distance between$G_{ijk}$ and $I_{ijk}$&$m$\\  
\it{$d_1$}&Distance between$I_{ijk}$ and t$M_ijk$&$m$\\  
\it{$d_2$}&Distance between$M_{ijk}$ (or $IP2$) point and $IP1$&$m$\\    
\it{$f$}&Free energy per unit volume&$J/m^3$\\
\it{$f_i$}&Immersed boundary force&$N$\\  
\it{$f_{bi}$}&Body forces&$N$\\  
\it{$G_{ijk}$}&Ghost point&$-$\\ 
\it{$I_{ijk}$}&Intersect point&$-$\\    
\it{$IP_{1}$}&First interpolation point&$-$\\ 
\it{$IP_{2}$}&Second interpolation point&$-$\\ 
\it{$l_s$}&Slip length&$m$\\
\it{$M$}&Mobility coefficient&$m^4/Ns$\\
\it{$M^*$}&Non-dimensional mobility coefficient&$-$\\  
\it{$M_{ijk}$}&Mirror point&$-$\\ 
\it{$n_i$}&Normal vector&$-$\\   
\it{$P$}&Pressure&$N/m^2$\\
\it{$Pe$}&P\'{e}clet number&$-$\\   
\it{$Re$}&Reynolds number&$-$\\  
\it{$S$}&Stabilisation parameter&$-$\\    
\it{$u_i$}&Face centred velocity vector&$m/s$\\
\it{$u_s$}&Slip velocity&$m/s$\\
\it{$u^*$}&First prediction velocity&$m/s$\\
\it{$u^{**}$}&Second prediction velocity&$m/s$\\    
\it{$\tilde{u}_i$}&Cell centred velocity vector&$m/s$\\
\it{$V_s$}&Non-dimensional contact line friction coefficient&$-$\\      
\it{W1}&Weight of the averaging points for $IP1$&$-$\\   
\it{W2}&Weight of the averaging points for $IP2$&$-$\\   
\it{$x_{im}$}&Coordinate of the mirror point&$m$\\  
\it{$\alpha$}&Fluid volume fraction&$-$\\ 
\it{$\alpha_s$}&Coefficient of the Helmholtz equation&$1/m^2$\\ 
\it{$\Gamma$}&Auxiliary variable for semi-implicit method&$1/m^2$\\ 
\it{$\Gamma_r$}&Contact angle relaxation time&$s$\\    
\it{$\gamma$}&The ratio $d_1/d_n$&$-$\\   
\it{$\Delta t_c$}&Convection time scale&$s$\\ 
\it{$\Delta t_{\nu}$}&Viscous time scale&$s$\\    
\it{$\Delta t_{\sigma}$}&Surface tension time scale&$s$\\   
\it{$\Delta t$}&Time step&$s$\\  
\it{$\epsilon$}&Interface thickness& $m$\\ 
\it{$\mu$}&Dynamic viscosity&$Ns/m^2$\\
\it{${\mu}_f$}&Contact line friction coefficient&$Ns/m^2$\\  
\it{$\theta_{eq}$}&Contact angle&$Degree$\\  
\it{$\rho$}&Mass density&$kg/m^3$\\
\it{$\sigma$}&Surface tension coefficient&$N/m$\\
\it{$\phi$}&Chemical potential per unit volume&$J/m^3$\\
\it{$ \mathscr{F}$}&Total free energy&$J$\\ 
 \end{longtable}
\end{center}

\section{Introduction}\label{sec:introduction}
Motion of  a three-phase contact line occurs in a variety of industrial fields from coating to energy conversion processes, nucleate boiling, droplet dynamics, two-phase flow in porous media, and microelectronics cooling, to name a few (\cite{Sui2013b,Sui2013a,Yarin2006}). Despite numerous studies  have been performed on the contact line motion, the underlying physics is still a matter of debate.
 The difficulty  in  studying the contact line movement originates in the so-called "contact line singularity" which was first discussed by \cite{Moffat1964} and 
  \cite{Huh1971}.  These authors showed that the fluid flow, close to the contact line, is in the Stokes regime and exhibits singularities in both the shear stress and the pressure \citep{Krechetnikov2019}. In general, three main solutions have been proposed to remove the singularities close to the moving contact line.  As the first solution, a slip velocity (Navier boundary condition) can be applied at the surface near the contact line \citep{Dussan1979}.  Modelling the dynamic contact angle and   the formation of a precursor film are the two other well-known solutions (\cite{Sui2013b}).

In order to model the moving contact line problem, different methods have been implemented for tracking the interface and reconstructing the contact line. \citep{Renardy2001} used a Volume of Fluid (VOF) method to model the contact line problem. A piecewise linear interface construction scheme was used to reconstruct the interface based on an indicator function.  \cite{AFKHAMI20095370} presented a mesh-dependent contact angle model to remove the stress singularity at the contact line.
\cite{Abhijit2007}  proposed a Level-Set approach to study bubble growth during an ebullition cycle. \cite{SPELT2005389} used an extended Level-set method to simulate multiple moving contact lines. The model accounts for flow inertia, contact line slip velocity and contact-line hysteresis. A front-tracking method was used by \cite{Muradoglu2010} to model the impact and spreading of viscous droplets on solid walls. \cite{PhysRevFluids.1.023302} studied the effects of viscoelasticity on drop impact and spreading on a flat solid surface using a front-tracking method together with finitely extensible nonlinear elastic-Chilcott-Rallison model for fluid elasticity. They showed that during the spreading phase, viscoelastic effects increase the spreading.  \cite{Afkhami} studied the role of surface wetting on interface instability and penetration modes in a porous medium consisting of two immiscible fluids. Their results suggest that the displacement patterns depend on both the capillary number and the surface wetting properties. They also examined the well-known Haines jumps (sudden interface jumps from one site to another) by analysing the characteristic time and length scales of the jumps.

 Diffuse interface models have also been used extensively to study the contact line dynamics.  Among different diffuse interface models, the phase-field method has drawn more attentions during the last decades. In a diffuse interface model, an order parameter (concentration) is defined to distinguish between different phases. The interface has a finite thickness within which the order parameter and the fluid properties vary smoothly (but significantly) from one phase to another one.  This assumption allows tracking the interface by solving an advection-diffusion equation for the order parameter. In this context, the Allen-Cahn model \citep{Allen1979}  is a reaction-diffusion phase-field equation which has been used to study phase separation in multi-component systems. Among others, \cite{Marouen2014} studied the equilibrium wetting  of a system of  immiscible fluids on a flat  substrate using the Allen-Cahn model.  As concerns dynamic wetting, however, the Cahn-Hilliard model \cite[see][]{Cahn1961} has been shown to be more reliable to remove the contact line singularity: imposing a dynamic contact angle boundary condition together with slip velocity at the wall is straight-forward in the Cahn-Hilliard formulation. Moreover, the Cahn-Hilliard equation
 provides the global conservation of the indicator function.  \cite{Jacqmin1999} introduced a dynamic boundary condition for the contact angle together with a model for the contact line slip velocity.  More recently, \cite{Carlson2011} discussed the importance of  a dynamic contact angle model for rapid wetting problems.  \{\cite{Ugis} employed molecular dynamics (MD) simulations together with the Cahn-Hilliard formulation of the phase-field model to study the contact line motion of water over a no-slip substrate. By comparing the results of the MD simulations and those of the phase-field model, these authors suggest that the phase-field mobility parameter and the local slip length are of great importance for the accuracy of the continuum model. 
From mathematical and numerical perspectives, solving  the Cahn-Hilliard equation (a non-linear fourth order partial differential equation), together with dynamic contact angle and slip velocity boundary conditions, is  a cumbersome task. The difficulty increases for complex wall geometries, especially if a body conformal mesh is used. Generating a high-quality body conformal numerical mesh on a complex geometry requires a significant effort. Moreover, a new mesh is needed to study a different solid substrate. An alternative to a body conformal numerical mesh are the immersed boundary methods \citep{Peskin2002}, a powerful tool to model fluid flows over complex geometries. This approach has been extensively  used to simulate fluid-solid interaction problems, mainly for a single  phase fluid. The idea of the immersed boundary method is to solve the system of equations on a cartesian numerical mesh (regardless of the solid geometry), and imposing the boundary conditions by adding  forces at specific grid points close to the boundary.

During the last decades, immersed boundary methods have been adopted with different interface tracking approaches, front-tracking \citep{DEEN20092186}, volume of fluid \citep{PATEL201728}, level set \citep{WANG201835}, phase field \citep{LIU2015484,Nishida2018}, with the aim to simulate the interaction between a multiphase fluid flow and
 a solid boundary. However, to the best of our knowledge, a fully Eulerian numerical approach for modelling the dynamic motion of a three phase contact line together with contact line slip velocity over any arbitrary geometry has not been reported yet. Since both the immersed boundary method and the phase field model proposed here are fully Eulerian, parallelisation of the algorithm is straight-forward and potentially highly scalable. The existing parallelisation of the underlying Navier-Stokes solver can be used for the multiphase part as well. 

The goal of this paper is to present a fully Eulerian modular hybrid algorithm for studying contact line motion on any arbitrary solid substrate. To properly model the contact line motion and remove the contact line singularity, we choose the Cahn-Hilliard phase-field formulation to track the interface; hence, an immersed boundary method is used together with the phase-field model with dynamic contact angle and slip velocity boundary conditions. The emphasis is on coupling the phase field model to the immersed boundary method.
 Depending on the problem under study and the available computational resources,  each of the modules of the proposed hybrid algorithm can be modified and extended independently. For instance, for simplicity of presentation, all the equations are solved explicitly in the manuscript. A more-efficient semi-implicit version of the algorithm is presented in Appendix A. Indeed, previous studies show that a semi-implicit implementation enables us to increase the numerical time step, increases the stability of the method, and reduces the numerical error \citep{DONG2012,SHEN2015,YU2017665,HUANG2020109192}. Here, we report numerical validations to show that the proposed algorithm provides accurate results for various test cases, for both choices of time integration (explicit and semi-implicit).\\  
 The outline of the manuscript is as follows. In section \ref{Sec:PFM}, we explain the main concept behind the phase field model together with the corresponding boundary conditions. In section \ref{sec:Numercis}, we summarise the implemented numerical schemes, whereas we elaborate on the proposed hybrid phase field-immersed boundary model in section \ref{Sec:recipie}. To validate and test our implementation, we report results from several numerical tests in section \ref{Sec:Tests}. We conclude our work in section \ref{sec:conclusion}, and finally present a more accurate semi-implicit version of the algorithm in Appendix A.

\section{Phase Field Model} \label{Sec:PFM}
Eulerian interface tracking approaches can be divided into two main groups, namely, sharp interface methods and diffuse interface methods.  In diffuse interface methods, the interface is assumed to have a finite thickness. Although the interface is much thicker than the real physical one, this assumption provides resolvable properties which vary continuously within the interface. Such a continuum model avoids any requirement for jump conditions at the interface or interface reconstruction. Moreover, fluid properties are conserved within the interface. During the last decades, the Phase Field Model (PFM) has become more and more popular in the multiphase flow community for these properties. It originates from Van der Waals model for free energy \citep{vanderWaals1893} where the bulk free energy and the interfacial free energy are added to give the total free energy per unit volume, $f$, of a system of two immiscible fluids as follows:
\begin{linenomath}\begin{equation} \label{FreeEnergy}
\begin{aligned}
\begin{gathered}
f= \frac{1}{2} \epsilon \sigma   \frac{\partial C}{\partial x_i}  \frac{\partial C}{\partial x_i} +\frac{\sigma}{\epsilon} \psi (C),
\end{gathered}
\end{aligned}
\end{equation}\end{linenomath}
\sloppy
where $C$ is the concentration (order parameter) which distinguishes different phases; it varies from $-1$ in one phase to $+1$ in the other. The variation of $C$ through the interface is smooth but rapid. $\psi$ is a double-well function, $\psi={(C^2-1)}^2/4$, with two minima for each stable phase. $\sigma$ and $\epsilon$ denote the surface tension coefficient between the two phases and the interface thickness, respectively. The first term in equation (\ref{FreeEnergy}) represents the contribution of the interfacial energy, whereas the second one models the bulk free energy density \citep{Jacqmin1999}.

Considering the requirement of minimum free energy in the equilibrium state, and defining the chemical potential $\phi$ as the variation of the total free energy ($ \mathscr{F}=\int{f}dV)$ with respect to the concentration, i.e.,~$\phi =\partial \mathscr{F}/\partial C$, Cahn and Hilliard proposed an equation for the evolution of the concentration where the motion of a diffuse interface within a binary fluid is modelled by the so-called Cahn-Hilliard equation \citep{CahnHilliard1958,Cahn1961}:
\begin{linenomath}\begin{equation} \label{Cahn-Hilliard}
\begin{aligned}
\begin{gathered}
\frac{\partial C}{\partial t} + u_i \frac{\partial C}{\partial x_i} = \frac{3}{2 \sqrt 2} \frac{\partial }{\partial x_i}  \left(   M  \frac{\partial \phi}{\partial x_i}  \right) ,\\
\end{gathered}
\end{aligned}
\end{equation}\end{linenomath}
where  $u_i$, and $M$ represent the fluid velocity vector and the mobility coefficient. The Cahn-Hilliard equation is an advection-diffusion equation which, in the limit of zero diffusivity, reduces to  a sharp interface model. Plenty of studies have considered the sharp interface limit, obtaining an appropriate range of values for the mobility coefficient based on the interface thickness \citep{Magaletti, Xu2018}.

The difference in the chemical potential between the two phases at the interface is the mechanism driving the interface motion (besides the advection of the interface by the mean flow). Theoretically, the chemical potential is the variation of the free energy with respect to the concentration and can be calculated with the following equation:
   \begin{linenomath}\begin{equation} \label{Chemical-Potential}
 \begin{aligned}
\begin{gathered}
 \phi =  \frac{\sigma}{\epsilon} {\psi}'(C) -\sigma \epsilon  \frac{\partial }{\partial x_i}  \left( \frac{\partial C}{\partial x_i} \right) .
 \end{gathered}
\end{aligned}
\end{equation}\end{linenomath}

To couple the Cahn-Hilliard equation (\ref{Cahn-Hilliard}) with the fluid flow, a term is added to the right hand side of the Navier-Stokes equation which accommodates for the surface tension forces at the interface \citep{Jacqmin1999}.
\begin{linenomath}\begin{equation} \label{NS equation}
\begin{gathered}
\frac{\partial (\rho u_i)}{\partial t} + \frac{\partial}{\partial x_j}(\rho u_i u_j) = -\frac{\partial P}{\partial x_i} + \frac{\partial}{\partial x_j} \left( \mu ( \frac{\partial u_i}{\partial x_j}+ \frac{\partial u_j}{\partial x_i})\right) + \phi \frac{\partial C}{\partial x_i}  ,\\
\frac{\partial (\rho u_i)}{\partial x_i} = 0 ,
\end{gathered}
\end{equation}\end{linenomath}
where $\rho$ and $\mu$ are the density and dynamic viscosity of the fluid, varying from $\rho_1$ and $\mu_1$ in one phase to $\rho_2$ and $\mu_2$ in the other one, defined as
\[\rho(C)= \left[ (C+1)\rho_2-(C-1)\rho_1  \right]/2\] and  \[ \mu(C)= \left[ (C+1)\mu_2-(C-1)\mu_1 \right]/2;\] $P$ is the pressure and $\phi \partial C/ \partial x_i$ represents the surface tension force at the interface. The second equation represents mass conservation for incompressible fluids.

In the presence of a solid substrate, a third term is added to eq.  (\ref{FreeEnergy}), the contribution of the solid substrate to the total free energy of the system \citep{ Carlson2011}:
\begin{linenomath}\begin{equation} \label{FreeEnergySolid}
\begin{aligned}
\begin{gathered}
\mathscr{F}= \int \left( \frac{1}{2} \epsilon \sigma   \frac{\partial C}{\partial x_i}  \frac{\partial C}{\partial x_i} +\frac{\sigma}{\epsilon} \psi (C) \right) dV+ \int \left(\sigma_{sg}+ (\sigma_{sl}-\sigma_{sg})\right)g(C)d\Gamma,
\end{gathered}
\end{aligned}
\end{equation}\end{linenomath}
where $\sigma_{sg}$ and $\sigma_{sl}$ are the surface tension coefficients between solid-gas and solid-liquid, respectively.
\begin{figure} [H]
    \begin{center}
     \includegraphics[width=0.7\textwidth]{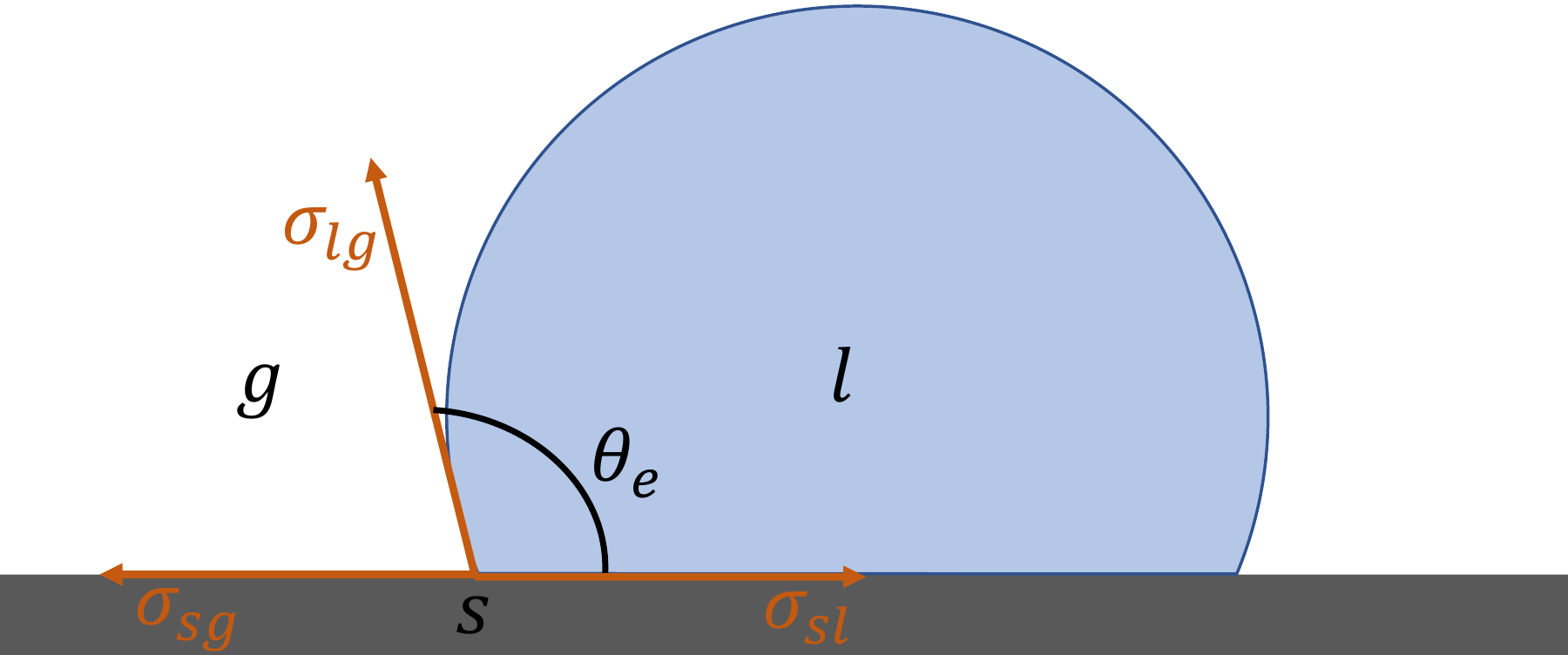}
\caption{Sketch of contact angle. According to Young's equation \citep{Young}, the equilibrium contact angle can be expressed as function of the surface tension coefficients between the different phases.}
\label{fig:YongEquation}
     \end{center}
     \end{figure}
According to Young's equation \citep{Young} , the equilibrium contact angle $\theta_e$ depends on the surface tension coefficient between each pair of the three phases, $ \sigma$, $\sigma_{sl}$, and $\sigma_{sg}$ (see figure \ref{fig:YongEquation}) as \[ \sigma_{sg}-\sigma_{sl}-\cos(\theta_{e})\sigma=0. \] The boundary condition for the concentration 
necessary to impose a prescribed contact angle is therefore obtained by minimising the energy at the solid wall \citep{Jacqmin1999,Jacqmin2000},
\begin{linenomath}\begin{equation} \label{WallBCS}  
\begin{gathered}
  \mu_f \epsilon \left( \frac{\partial C}{\partial t}+ u_i \frac{\partial C}{\partial x_i} \right)= \sigma \epsilon \frac{3}{2 \sqrt 2}\frac{\partial C}{\partial x_i}n_i + \sigma cos(\theta_{eq}) g'(C),\\
  \frac{\partial \phi}{\partial x_i}n_i =0,
   \end{gathered}
\end{equation}\end{linenomath}
where $\mu_f$, $\theta_{eq}$, and $n_i$ are the contact line friction coefficient, the equilibrium contact angle, and the vector normal to the solid surface. The first boundary condition in eq. (\ref{WallBCS}) models the dynamics of the contact line motion, where the contact line friction coefficient is inversely proportional to the time needed by the contact line to relax to its prescribed static contact angle \citep{ Carlson2012,Xu2018}. In this equation, $g(C)=({2}+{3}{C}-{C}^3)/4 $ is a function which varies smoothly between zero and 1 from one stable phase to the other. The second boundary condition in equation (\ref{WallBCS}) guarantees impermeability at the wall.
  
Finally, in the case of non-zero wall slip velocity, the following  equation can be solved together with the other boundary conditions to obtain the  slip velocity at the wall \citep{ Carlson2012}:
\begin{linenomath}\begin{equation} \label{SlipVelocity}
\begin{gathered}
 \frac{\mu}{l_{s}}u_{j_{s}}t_j= \mu \frac{\partial (u_j  t_j)}{\partial (x_in_i)}- \left[ \frac{3}{2 \sqrt 2}\frac{\partial C}{\partial x_i}n_i + \sigma cos(\theta_{eq}) g'(C)\right]\frac{\partial C}{\partial x_j}t_j ,\\
 \end{gathered}
\end{equation}\end{linenomath}
where $u_{s}$, $l_{s}$, and $t_j $ are the slip velocity, slip length, and the unit vector tangent to the surface. 

\section{Numerical method} \label{sec:Numercis}
The full system of equations introduced above can be summarised as follows:
\begin{linenomath}\begin{equation} \label{SystemOfEquation}
\begin{gathered}
\frac{\partial (\rho u_i)}{\partial t} + \frac{\partial}{\partial x_j}(\rho u_i u_j) = -\frac{\partial P}{\partial x_i} + \frac{\partial}{\partial x_j} \left( \mu ( \frac{\partial u_i}{\partial x_j}+ \frac{\partial u_j}{\partial x_i})\right) + \phi \frac{\partial C}{\partial x_i} + f_i + f_{b_i},\\
\frac{\partial (\rho u_i)}{\partial x_i} = 0 ,\\
\frac{\partial C}{\partial t} + u_i \frac{\partial C}{\partial x_i} = \frac{3}{2 \sqrt 2} \frac{\partial }{\partial x_i}  \left(   M  \frac{\partial \phi}{\partial x_i}  \right) ,\\
 \phi =  \frac{\sigma}{\epsilon} {\psi}'(C) -\sigma \epsilon  \frac{\partial }{\partial x_i}  \left( \frac{\partial C}{\partial x_i} \right),
\end{gathered}
\end{equation}\end{linenomath}
where $f_i$ represents the immersed boundary force used to account for the complex wall geometry, explained in section \ref{Sec:IBM}, and $f_{b_i}$ indicates all the other body forces (e.g. the gravitational force). The complete  boundary conditions at the wall are
\begin{linenomath}\begin{equation} \label{AllBCs}  
\begin{gathered}
  \mu_f \epsilon \left( \frac{\partial C}{\partial t}+ u_i \frac{\partial C}{\partial x_i} \right)= \sigma \epsilon \frac{3}{2 \sqrt 2}\frac{\partial C}{\partial x_i}n_i + \sigma cos(\theta_{eq}) g'(C),\\
  \frac{\partial \phi}{\partial x_i}n_i =0,\\
   \frac{\mu}{l_{s}}u_{j_{s}}t_j= \mu \frac{\partial (u_j  t_j)}{\partial (x_in_i)}- \left[\frac{3}{2 \sqrt 2} \frac{\partial C}{\partial x_i}n_i + \sigma cos(\theta_{eq}) g'(C)\right]\frac{\partial C}{\partial x_j}t_j ,\\
    \frac{\partial P}{\partial x_i}n_i =0,\\
    u_in_i = 0
   \end{gathered}
\end{equation}\end{linenomath}
As mentioned above, in the following we will introduce the numerical algorithm assuming a fully explicit approach. However, this set of equations can also be solved semi-implicitly as discussed in Appendix A. Note that we use both the algorithms alternatively in the numerical tests discussed later on, the differences between the results being negligible once the step is chosen correctly. However, the semi-implicit algorithm allows, on average, a 10 times larger time step.

\subsection{Time integration and spatial discretisation} \label{Sec:discretization}
We solve the system of equations (\ref{SystemOfEquation} and \ref{AllBCs}) on a Cartesian mesh with a staggered arrangement, where the velocity components are defined at the faces and the pressure, the chemical potential and the order parameter are defined at the cell centres. The second-order finite difference scheme is used for spatial discretisation and the different terms are advanced in time explicitly using the Adams-Bashforth scheme. Finally, the fractional-step method for incompressible two-fluid systems is implemented as in \cite{DODD2014416}. The baseline solver has been extensively validated in the previous works \citep[see among others][]{rosti_ge_jain_dodd_brandt_2019,ROSTI20183,DeVita2020,rosti_izbassarov_tammisola_hormozi_brandt_2018}.

\subsection{Immersed boundary method} \label{Sec:IBM}
During the last decades, a variety of immersed boundary methods (IBM) have been used for modelling fluid-solid interactions with moving and fixed bodies (\cite{IBMreview}). 
In most of the IBM formulations, the solid boundary is represented by a set of Lagrangian points whose locations are tracked by solving an extra set of equations. An auxiliary force is added to the mesh cells surrounding each of the Lagrangian points to impose the no-slip and no-penetration conditions at the solid boundaries. However, there are also fully Eulerian IBM formulations where the immersed boundary forces are computed directly on the numerical grid points, especially for the case of fixed objects. 

In this paper, we use a simple Eulerian IBM formulation, namely, the volume penalisation model proposed by \cite{Kajishima2001} to impose the no-slip velocity boundary condition at and zero velocity inside the solid wall. For the slip velocity and the dynamic/static contact angle boundary conditions we use a ghost-cell approach, discussed in the next section. The Eulerian approach proposed here for the IBM formulation facilitates the implementation and especially the code parallelisation.

Let us define the fluid volume fraction at each grid cell as the ratio of the volume of the cell which is occupied by the fluid to the total cell volume and denote it  by $\alpha_{ijk}$ with $(i,j,k)$ the cell index. The volume fraction varies between zero (for a cell entirely located in the solid) to one (for a cell entirely located in the fluid). Kajisjima et al. suggested to calculate the IBM force, $f_i$, and to modify the prediction velocity, $u_{ijk}^*$, obtained by integrating in time the momentum equations under the action of viscous stresses and surface tension only, as follows
\begin{linenomath}\begin{equation} \label{Penalization1}
\begin{gathered}
f_{ijk} = (1-\alpha_{ijk}) \frac{{\left(u_{sol}-u^*\right)}_{ijk}}{\Delta t},\\
u_{ijk}^{**} =  u_{ijk}^{*} + \Delta t f_{ijk},
 \end{gathered}
\end{equation}\end{linenomath}
 where $u_{ijk}^{**}$ is the second prediction velocity and $u_{sol}$ is the solid wall velocity within the corresponding grid cell. In the case of stationary wall ($u_{sol}=0$), equation (\ref{Penalization1}) reduces to the following simpler form,
\begin{linenomath}\begin{equation} \label{Penalization2}
\begin{gathered}
u_{ijk}^{**} =  u_{ijk}^{*}  \alpha_{ijk}.
 \end{gathered}
\end{equation}\end{linenomath}

%In this paper, as it was mentioned before,  a staggered grid is used to solve the system of equations. Therefore, for a given grid point whose index is  $(i,j,k)$, volume fractions should be calculated for four different cells (three cells for face-centred properties and one cell for cell-centred properties). It is worth to mention that although this computation would be expensive, volume fractions are calculated only once for any simulation.

\section{The hybrid PFM-IBM Algorithm } \label{Sec:recipie}
To impose the boundary conditions for the order parameter and slip velocity at the wall, we need to calculate the values of the different quantities appearing in equations (\ref{AllBCs}) (the order parameter, velocity components, derivatives of the order parameter, etc.) at the wall, separating  tangential and normal components
with respect to the wall surface. To accomplish this, we need to pre-compute some different quantities, e.g.\ solid volume fraction, normal vector, averaging weights, etc. Having performed this initialisation step, it is possible to integrate in time the governing equations
with general boundary conditions at the wall. We now proceed to provide a numerical recipe to solve the system at hand  with corresponding boundary conditions for slip velocity and order parameter at the wall  for a two-dimensional system. The method can be easily extended to three-dimensional problems, as shown in the result section.

\subsection{Preliminary computations}
In this section, we elaborate the details of the preliminary calculations needed for the proposed algorithm (green dashed box in figure \ref{fig:Chart}).

\subsubsection{Volume fraction and normal vectors} \label{Sec:NormalVect}
In order to impose all the boundary conditions introduced in equations \ref{WallBCS} and \ref{SlipVelocity} together with the volume penalisation immersed boundary method (equation \ref{Penalization2}), we should first calculate the liquid and solid volume fraction of each grid cell, $\alpha_{ijk} $. This needs to be done for all the four numerical cells on a staggered-grid, namely, one cell-centred cell and three face-centred cells for each grid point $ijk$.

\begin{figure} [H]
    \begin{center}
     \includegraphics[width=0.5\textwidth]{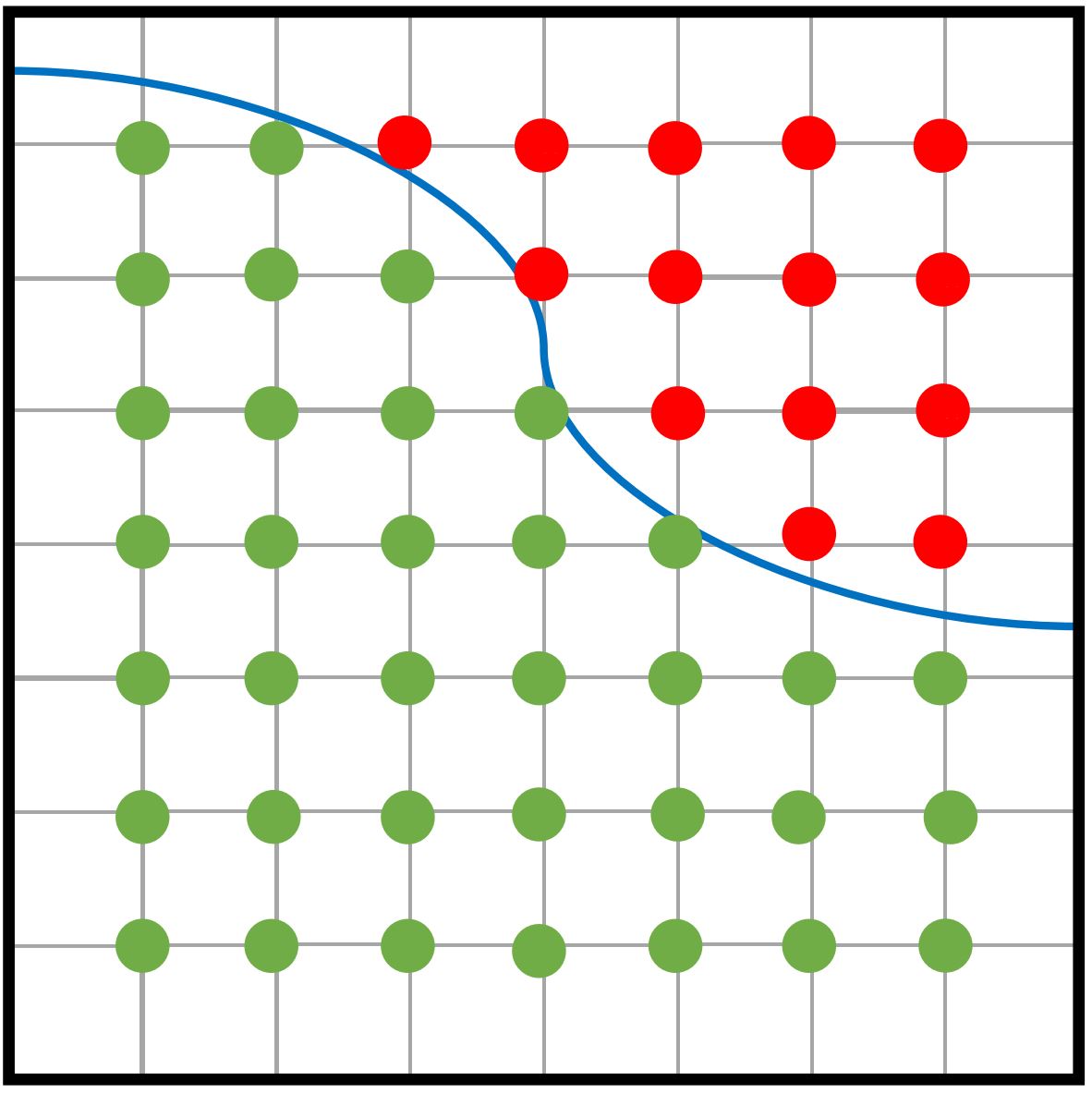}
\caption{Computing the volume fraction of each cell. Blue line represents the solid wall border, green dots are the subgrid points located inside the solid and the red dots are those in the fluid phases.}
\label{fig:VolumeFraction}
     \end{center}
     \end{figure}

We propose a simple approach for computing the volume fractions. Let us consider a cell-centred cell at the grid point $(i,j,k)$ and, first, divide the cell into a sufficient number of subgrid points in both directions as shown in figure \ref{fig:VolumeFraction} (our tests suggest that 100 points in each direction are enough). Under the assumption that the coordinates of each subgrid point in the cell are known, we can determine whether the point is inside or outside the solid wall. Hence, the volume fraction $\alpha$ is simply the ratio of the number of subgrid points outside the solid to the total number  in the cell.
Having calculated the volume fraction on each cell, the solid wall normal vectors are approximated with the gradient of the volume fraction. To this aim, we first compute the  derivative of the volume fraction at the four cell corners \citep{Satoshi2012}, e.g.,
\begin{linenomath}\begin{equation} \label{NormalVectors_part1}
\begin{gathered}
m_{x_{1_{i+1/2,j+1/2}}} = \frac{\alpha_{i+1,j}+\alpha_{i+1,j+1}-\alpha_{i,j}-\alpha_{i,j+1}}{\Delta x_{1_i}+ \Delta x_{1_{i+1}}},\\
m_{x_{2_{i+1/2,j+1/2}}} = \frac{\alpha_{i,j+1}+\alpha_{i+1,j+1}-\alpha_{i,j}-\alpha_{i+1,j}}{\Delta x_{2_i}+ \Delta x_{2_{j+1}}},\\
 \end{gathered}
\end{equation}\end{linenomath}
The value of the derivatives can be calculated in the three other corners in a similar fashion and then be used altogether to compute the normal vector at the cell center as
\begin{linenomath}\begin{equation} \label{NormalVectors_part2}
\begin{gathered}
n_{x_{1_{i,j}}} = \frac{ \Delta x_{1_i} m_{x_{1_{ij}}}  } {\sqrt{   {(m_{x_{1_{ij}}}   ) }^2 + {(m_{x_{1_{ij}}}   ) }^2+\epsilon_0}},\\
n_{x_{2_{i,j}}} = \frac{ \Delta x_{2_i} m_{x_{1_{ij}}}  } {\sqrt{   {(m_{x_{1_{ij}}}   ) }^2 + {(m_{x_{2_{ij}}}   ) }^2+\epsilon_0}},
 \end{gathered}
\end{equation}\end{linenomath}
where $\epsilon_0$ is a very small positive number used to avoid division by zero and 
\begin{linenomath}\begin{equation} 
\begin{gathered}
 m_{x_{1_{ij}}} = \frac{m_{x_{1_{i+1/2,j+1/2}}}+m_{x_{1_{i+1/2,j-1/2}}}+m_{x_{1_{i-1/2,j+1/2}}}+m_{x_{1_{i-1/2,j-1/2}}}}{4},\\
 m_{x_{2_{ij}}} = \frac{m_{x_{2_{i+1/2,j+1/2}}}+m_{x_{2_{i+1/2,j-1/2}}}+m_{x_{2_{i-1/2,j+1/2}}}+m_{x_{2_{i-1/2,j-1/2}}}}{4}.
 \end{gathered}
\end{equation}\end{linenomath}

\subsubsection{Ghost points, intersections, mirror and interpolation points} \label{Sec:GhostIntersectMirror}
Next, we need to identify four groups of points: the ghost, intersect, mirror, and interpolation points. In this section we use the term \textit{temporary array} to refer to lists that are defined and used only in the initialisation steps and can be deallocated later on. On the other hand, the term \textit{permanent array} refers to lists that contain variables required during the whole simulation.

\begin{figure} [H]
\begin{center}
 \includegraphics[width=0.5\textwidth]{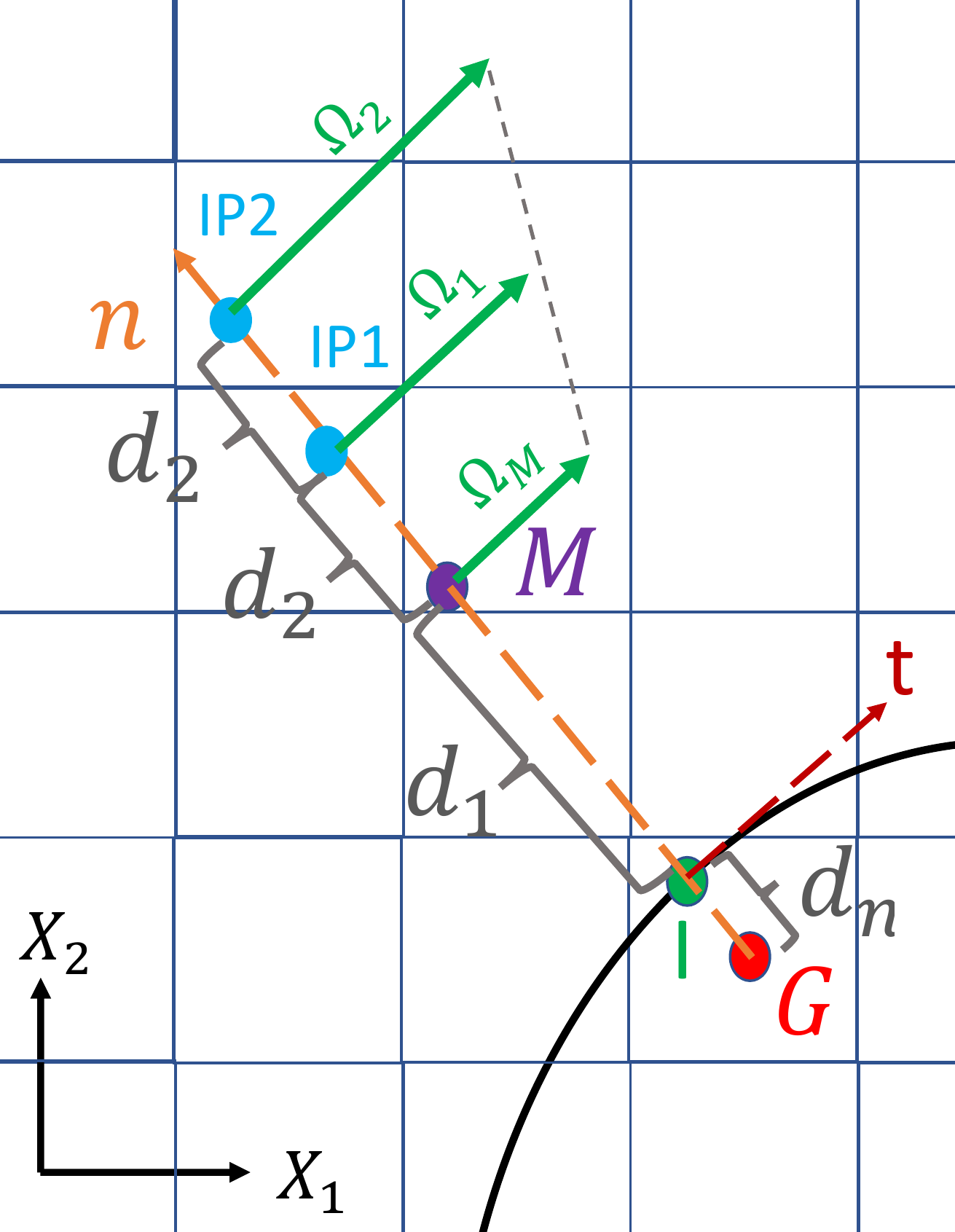}
\caption{Schematic of the IBM treatment and interpolation procedure.}
\label{fig:IBM}
 \end{center}
 \end{figure}

According to figure \ref{fig:IBM}, we define the ghost points ($G_{ijk}$) as the cell-centred points inside the solid having at least one neighbour in the fluid phase. In addition to the ghost points, we need to find the intersect points: starting from the ghost point, we march along the normal direction towards the fluid phase with a small enough step size. At each marching step, we verify whether the new point is inside or outside the solid; the first point outside the solid is labeled as the intersect point, $I_{ijk} $ corresponding to the ghost point $G_{ijk}$. The distance $d_n$ between each ghost point and the corresponding intersect point is saved in a permanent array for later use.

For each ghost point $G_{ijk}$ we also identify a mirror point $M_{ijk}$, such that the ghost point, intersect point, and the mirror point are all aligned on a straight line normal to the wall; the distance between the intersect point and the mirror point ($d_1$) is the same for all the ghost points and is chosen long enough to ensure that the mirror point $M_{ijk}$ and the corresponding ghost point $G_{ijk}$ are not located in the same numerical cell ($d_1>\lvert dx_i \lvert/2$, $\lvert dx_i \lvert$ being the length of a cell diagonal). The coordinates are saved in a three dimensional temporary array whose index $(i,j,k)$ refer to the ghost point location. For instance, ${x_1}_{m}(i,j,k), {x_2}_{m}(i,j,k),$ and ${x_3}_{m}(i,j,k)$ represent the coordinate of the mirror point corresponding to the ghost point with index $i,j,k$.
%  For the interpolation purposes, we should identify the cells at which the mirror points are located. Again, we define three temporary arrays, namely $i_m(i,j,k)$, $j_m(i,j,k)$, $k_m (i,j,k)$. For a given ghost cell (given $ i,j,$ and $ k)$, we can detect $ i,j,$ and $k$ of the cell at which the corresponding mirror point is located.\\

Finally, we identify two additional sets of points, the interpolation points denoted $IP_1$ and $IP_2$. These points will be used to extrapolate the magnitude of any quantity at the mirror point as discussed later. The interpolation points are located on the same straight line as  the mirror, intersect, and ghost points. 
Note that the distance between the two interpolation points and between the first interpolation point and the mirror one is the same (i.e.\ $d_2$ in figure \ref{fig:IBM}). 
The magnitude of $d_2$ is chosen smaller that $d_1$, as function of the Reynolds number of the problem under investigation. As explained later, the magnitude of any arbitrary parameter at the mirror point is linearly extrapolated from the magnitudes at the corresponding interpolation points. This assumptions is valid if the interpolation points and the mirror points are located inside the inner-part of a boundary layer where a linear profile can be safely assumed. 
In general, it is well known that as the Reynolds number increases, the boundary layer becomes thinner; thus the numerical grid should be finer to resolve the boundary layer properly (regardless of the IBM algorithm). Nevertheless, it is important to check that the interpolation points and the mirror points are located inside the inner part of the boundary layer %However, it is important to check that all the IBM points are placed within the inner part of the boundary layer 
(and if not, tune the value of $d_2$).
%In this paper, we use the algorithm for flows at moderately low Reynolds number ($Re<10$) and therefore, considering the thickness of the inner-layer and the grid resolutoin, we make sure that the mirror points and the two interpolation points are within the inner-layer.
Note, finally, that we also need to identify the cells in which the interpolation points are located and store their indices in a permanent array.

\subsubsection{Coordinate transformation} \label{Sec:CoordinateTransformation}
Although we solve the governing equations using a simple Cartesian coordinate system ($X_1, X_2$ for two-dimensional problems), the phase field boundary conditions are expressed in a coordinate system following the solid boundary, with normal and tangential vectors $n_j$ and $t_j$. Therefore, we need to transform between the two systems using the calculated normal vector, $n_i$. For an arbitrary variable $\Omega$, subject to a coordinate transformation, we have
 \begin{linenomath}\begin{equation} \label{CoordinateTransformation}
\begin{gathered}
\Omega_n = \Omega_{X_1} n_1+ \Omega_{X_2} n_2  , \quad  \quad  \Omega_t = \Omega_{X_1} n_2-\Omega_{X_2} n_1,  \\
 \Omega_{X_1} = \Omega_n n_1+ \Omega_t n_2  ,\quad \quad  \Omega_{X_2} = \Omega_n n_2+ \Omega_t n_1,
 \end{gathered}
\end{equation}\end{linenomath}
where the suffixes $n$ and $t$ indicate the components in the frame following the boundary.

\subsubsection{Inverse distance weighting averaging and linear extrapolation} \label{InvDisAve}
Let us consider any arbitrary variable, say $\Omega$, whose approximated value is needed at point M. 
As first step, we average the value of $\Omega$ at the 2 interpolation points ($IP_1$ and $IP_2$ in figure \ref{fig:IBM}).  To do so,
 we identify from the coordinates of each of the interpolation points the grid points surrounding each of them ($AP_1$ to $AP_5$ in figure \ref{fig:IDW}).
 We then perform the averaging through an inverse distance weighting: for a two-dimensional problem, the average uses five points ($AP_1, AP_2, AP_3$, $AP_4$ and $AP_5$), with % providing an accurate enough value at the interpolation points. %However, for three-dimensional problems, we use seven interpolation points.
weights equal to the inverse of the squared distance between the  averaging points ($AP_i$) and the interpolation point ($IP$), denoted here $1/h_i^2$.   The interpolated value can thus be calculated as follows:
\begin{figure}[ H]
\begin{center}
\includegraphics[width=0.5\textwidth]{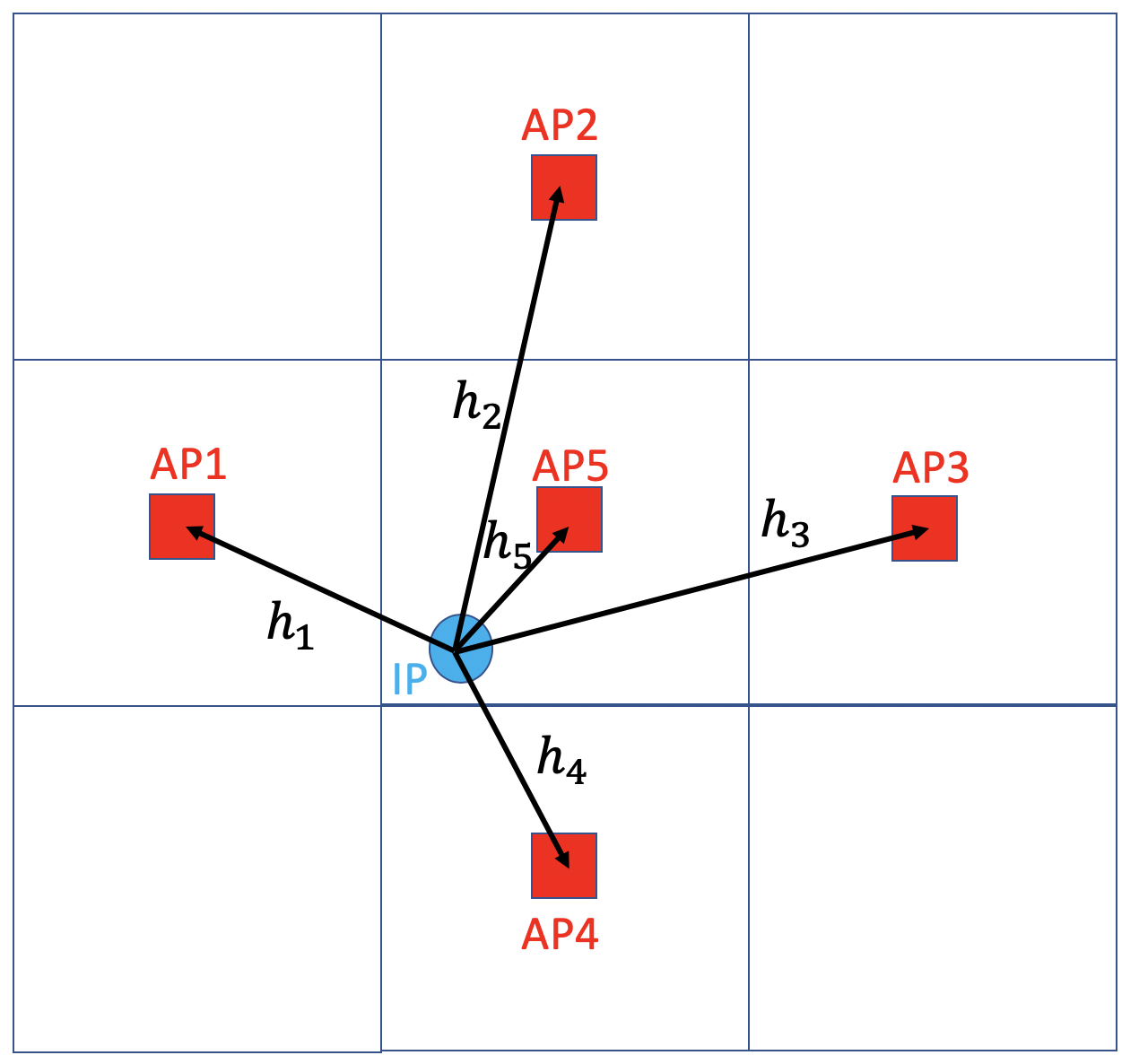}
\caption{Schematics of the inverse distance weighting averaging method used to obtain the values of a field variable $\Omega$  in $IP$ from 5 surrounding points.}
\label{fig:IDW}
\end{center}
\end{figure}
 \begin{linenomath}\begin{equation} \label{weright averaged} 
\begin{gathered}
\Omega_{IP} = \frac{1}{q} \sum_{i=1}^5 \omega(i)\Omega(i)\\
\omega(i) =  \left(\frac{1}{h_i}\right)^2, \quad q = \sum_{i=1}^5 w_i
\end{gathered}
\end{equation}\end{linenomath}
where $w(i)$ is the weight of the $i^{th}$ averaging point and $q$ is the sum of all the weights. Depending on the geometry, the interpolation point can overlap with one of the averaging points, in which case $h_i$ goes to zero and  the interpolated value  can be taken as the value  at the averaging point. To carry out the computation of the boudnary condition, we therefore define two other permanent arrays, $W1(i,j,k,AP)$ and $W2(i,j,k,AP)$: $W1(i,j,k,AP)$ contains the weight of each averaging point ($AP1$ to $AP5$) corresponding to the first interpolation point $IP1$ of any ghost point $G_{ijk}$, and similarly for $W2(i,j,k,AP)$ containing the weights for the second interpolation point  $IP2$.

%After computing and saving the above mentioned weight arrays, we deallocate all the temporary arrays we have defined before. 
At the end of this preliminary phase, we are ready to use, during the simulation, the weight arrays to compute the averaged value of the variable $\Omega$ at each of the interpolation points and then use these to find the value at the mirror point by linear extrapolation
\begin{linenomath}\begin{equation} \label{extrapolation}
\Omega_M = 2\Omega_{IP1}-\Omega_{IP2}.
\end{equation}\end{linenomath}

\subsection{Solution algorithm}
In this section we describe the solution of the system of equations and the algorithm used to impose the boundary conditions on a fixed wall of arbitrary shape.

\subsubsection{Solving the Cahn-Hilliard equation} \label{SolviingCahnHilliard}
We first solve the Cahn-Hilliard equation to update the order parameter from time $C^n$ to $C^{n+1}$. Note that irrespective of the time integration method (explicit or semi-implicit), by solving the Cahn-Hilliard equation, we update the correct value of the order parameter in all the numerical grid points (even inside the solid) except for the ghost points. The details of the semi-implicit algorithm adopted for the Cahn-Hilliard equation are presented in appendix A. For the explicit algorithm, we use the second-order central finite difference scheme for the spatial discretisation and  second order Adam-Bashforth for the temporal discretisation.
\begin{linenomath}\begin{equation} \label{ExpCH} 
\begin{gathered}
\frac{C^{n+1}-C^n}{\Delta t} = \frac{3}{2} \left( -u_i^n \frac{\partial C^n}{\partial x_i}\right)+\frac{1}{2} \left(u_i^{n-1} \frac{\partial C^{n-1}}{\partial x_i}\right)\\
+\frac{3}{2}\left[ \frac{3}{2\sqrt{2}} \frac{\partial}{\partial x_i} \left( M\frac{\partial \phi^n}{\partial x _i} \right) \right]-\frac{1}{2}\left[ \frac{3}{2\sqrt{2}} \frac{\partial}{\partial x_i} \left( M\frac{\partial \phi^{n-1}}{\partial x _i} \right) \right].
\end{gathered}
\end{equation}\end{linenomath}
 
\subsubsection{Imposing the boundary conditions for the order parameter}

Of relevance here, we impose the contact angle and the no mass penetration boundary conditions at the ghost points using the IBM algorithm. %Therefore, we simply impose Neuman boundary conditions in the $X2$ direction and periodic boundary condition in $X1$ direction of the computational domain boundary for order paramter and chemical ponetial (CD in figure \ref{fig:Chart}).\\
According to equation (\ref{WallBCS}), the boundary conditions for the order parameter are defined based on the fluid properties and their derivatives at the wall. Particularly, we need the value of the order parameter, $C$, and its derivatives in the normal and the tangential directions ($\partial C/ \partial n$,  $\partial C/\partial {x_t}$), and the wall-normal and tangential velocity components at the wall  ($V_n$,$V_t$). 
Note that, we denote the derivative in the normal direction by $\partial / \partial n$, while in tangential direction by  $\partial / \partial x_t$ to avoid confusion with the time derivative $ \partial / \partial t$. The aforementioned values at the wall are found by using the properties of the fluid at the ghost points and the interpolated values at the mirror points together with the coordinate transformation from ($X_1, X_2$) to ($n,t$). The boundary conditions (equation \ref{WallBCS}) are solved at the cell centre; therefore, all the velocity components are first interpolated at the cell centre (here denoted by a tilde). 

To impose the boundary conditions, we proceed as follows. Let us define $\gamma$ as the ratio of the distance between the ghost and the intersect points to the distance between the intersect point and the mirror point ($\gamma= d_1/dn$). The value of the order parameter, the derivatives of the order parameter, and the velocity components at the intersect point can be calculated as
\begin{linenomath}\begin{equation} \label{interpolationAtWall}
\begin{gathered}
{C_I}^{n+1} = \frac{ \gamma \hat{C}_G +C^{n+1}_M}{\gamma+1 },\\
\left(\frac{\partial C}{ \partial X_1}\right)^{n+1}_I = \frac{\gamma \left(\frac{\partial C}{ \partial X_1}\right)^{n+1}_G +\left(\frac{\partial C}{ \partial X_1}\right)^{n+1}_M}{\gamma+1}, \\
\left(\frac{\partial C}{ \partial X_2}\right)^{n+1}_I = \frac{\gamma \left(\frac{\partial C}{ \partial X_2}\right)^{n+1}_G +\left(\frac{\partial C}{ \partial X_2}\right)_M^{n+1}}{\gamma+1}, \\
{\tilde{u}}^{n}_{1_I} = \frac{\gamma {\tilde{u}^{n}_{1_G} +\tilde{u}^{n}_{1_M}}}{\gamma+1}, \\
{\tilde{u}}^{n}_{2_I} = \frac{\gamma {\tilde{u}^{n}_{2_G} +\tilde{u}^{n}_{2_M}}}{\gamma+1}.
\end{gathered}
\end{equation}\end{linenomath}
As already mentioned, we do not update the magnitude of the order parameter at the ghost point when solving the Cahn-Hilliard equation. Therefore, we use here $\hat{C}_G$ which is an estimation of the order parameter at the ghost point at time $n+1$, defined as $\hat{C}_G= 2C^n_G-C^{n-1}_G$. The different quantities  are then projected from $X_1, X_2$ to $n, t$
\begin{linenomath}\begin{equation} \label{ProjectingToNT}
\begin{gathered}
{v_t}^{n}_I = {u_1}^{n}_I n_2-{u_2}^{n}_I n_1, \\
%{v_t}^{n}_G = {u_1}^{n}_G n_2-{u_2}^{n}_G n_1, \\
%{v_t}^{n}_M = {u_1}^{n}_M n_2-{u_2}^{n}_M n_1, \\
\left( \frac{\partial C}{\partial x_j}t_j\right)^{n+1}_I = \left( \frac{\partial C}{\partial X_1}\right)^{n+1}_I n_2-\left( \frac{\partial C}{\partial X_2}\right)^{n+1}_I n_1, \\
\left( \frac{\partial C}{\partial n}\right)^{n+1}_I = \frac{C^{n+1}_M+(\gamma^2-1)C^{n+1}_I-\gamma^2 \hat{C}_G}{\gamma (\gamma-1)dn}.
\end{gathered}
\end{equation}
Finally, we can compute $\partial C / \partial t$ at the wall as
\begin{equation} \label{WallBCS_recast}  
\begin{gathered}
    \left(\frac{\partial C}{\partial t}\right)_I^{n+1}  = -{v_t}^{n}_I \left(\frac{\partial C}{\partial x_j}\right)^{n+1}_I + \left( \frac{1}{\mu_f \epsilon} \right)\left[ \sigma \epsilon \frac{3}{2 \sqrt 2}\left(\frac{\partial C}{\partial n }\right)^{n+1}_I + \sigma cos(\theta_{eq}) g'(C_I^{n+1})\right].
   \end{gathered}
\end{equation}\end{linenomath}
By integrating in time, we find the updated value of the order parameter at the wall $C_I^{(n+1)}$ which is used to update the order parameter at the corresponding 
ghost point as
\begin{linenomath}\begin{equation} \label{UpdateC}
\begin{gathered}
C^{n+1}_G = \frac{(\gamma+1)C^{n+1}_I -C^{n+1}_M}{\gamma}.
\end{gathered}
\end{equation}\end{linenomath}
We can now update the chemical potential and impose the corresponding boundary condition ($\partial \phi/ \partial n = 0$) in a similar way. Finally, the density and viscosity are updated using the new values of $C$.

\subsubsection{Calculating the first and the second prediction velocities}
As illustrated in figure \ref{fig:Chart}, the next step is to solve the Navier-Stokes equations. Similar to the Cahn-Hilliard equation, we calculate the first and the second prediction velocities using either an explicit or semi-implicit algorithm. The details of the semi-implicit implementation are provided in appendix A. For the explicit algorithm, we use second order central finite differences for the spatial discretisation and we integrate all the terms in time using the second order Adam-Bashforth scheme. Note that we solve the Navier-Stokes equation for all the numerical grid points except at the ghost points, which  we will use to impose the boundary conditions.
\begin{linenomath}\begin{equation} \label{PredictionExp}  
\begin{gathered}
\frac{u^*_i-u^n_i}{\Delta t} = -\left( \frac{3}{2} \frac{\partial}{\partial x_j}(u^n_i u^{n}_j)-\frac{1}{2} \frac{\partial}{\partial x_j}(u^{n-1}_i u^{n-1}_j) \right)+
\left(  \frac{3}{2} \frac{\phi^{n+1}}{\rho^{n+1}}\frac{\partial C^{n+1} }{\partial x_i}- \frac{1}{2} \frac{\phi^{n+1}}{\rho^{n+1}}\frac{\partial C^{n+1} }{\partial x_i}    \right)+\\
\left[  \frac{3}{2}  \frac{\partial}{\partial x_j} \left(\mu^{n+1}(\frac{\partial u^n_i}{\partial x_j}+\frac{\partial u^n_j}{\partial x_i})\right) -\frac{1}{2}  \frac{\partial}{\partial x_j} \left(\mu^{n+1}(\frac{\partial u^n_i}{\partial x_j}+\frac{\partial u^n_j}{\partial x_i})\right)  \right]+
\left( \frac{3}{2}  f^n_{b_i}-\frac{1}{2}  f^{n-1}_{b_i} \right),\\
u^{**}_i = u^{*}_i \alpha
\end{gathered}
\end{equation}\end{linenomath}
We recall that  $f_{b_i}$ is the summation of all the external body forces (such as gravity). The last equation, the step between $u_i^*$ and $u_i^{**}$, is the
IBM penalisation discussed above, which imposes zero velocity inside the solid. In the cases with slip, the boundary condition at the wall is modified using the ghost point, so that the fluid has a slip velocity. This is detailed in the next section.  

\subsection{Enforcing the velocity boundary conditions}
We impose the slip velocity boundary condition at the ghost point using the IBM algorithm. %Therefore, we simply impose no-slip and no penetration boundary conditions in the $X2$ direction and periodic boundary condition in $X1$  direction of the computational domain boundary (CD in figure \ref{fig:Chart}).\\ 
To impose the velocity boundary conditions, we first interpolate the calculated second prediction velocity $u^{**}$ at the cell centres. Next, we use the interpolation scheme 
introduced in section \ref{InvDisAve} to calculate the magnitude of any cell-centred velocity component at the mirror points. Due to the no penetration boundary condition at the wall, the normal component of the velocity at the wall is equal to zero. Therefore, the normal component of the velocity at the ghost point can be updated as
  \begin{linenomath}\begin{equation} \label{NoPenetration}
   V^{**}_{n_G} =  -\frac{V^{**}_{n_M}}{\gamma}.
  \end{equation}
The boundary conditions for the tangential component of the velocity at the wall can be obtained from equation (\ref{SlipVelocity})
\begin{equation} \label{SlipVelocityTangen}
\begin{gathered}
 \frac{\mu}{l_{s}}V^{**}_{t_{s}}= \mu \frac{\partial V^{**}_t}{\partial n}- \left[ \frac{\partial C^{n+1}}{\partial n} + \sigma cos(\theta_{eq}) g'(C^{n+1})\right]\frac{\partial C^{n+1}}{\partial x_t}. \\
 \end{gathered}
\end{equation}\end{linenomath}
After discretizing the equation, we can obtain the tangential velocity at the ghost point by solving the following equation:

\begin{linenomath}\begin{equation} \label{UpdateVtan}
\begin{gathered}
 \frac{\mu}{l_{s}}\frac{\gamma V^{**}_{t_G}+V^{**}_{t_M}}{\gamma+1}= \mu \frac{V^{**}_{t_M}+(\gamma^2-1)V^{**}_{t_I} -\gamma^2 V^{**}_{t_G}}{\gamma(\gamma+1)dn}- \left[ \frac{\partial C^{n+1}}{\partial n} + \sigma cos(\theta_{eq}) g'(C^{n+1})\right]\frac{\partial C^{n+1}}{\partial x_t} \\
 V^{**}_{t_G} = \frac{ V^{**}_{t_m} (l_s-dn)}{l_s+\gamma dn}-  \left[ \frac{\partial C^{n+1}}{\partial n} + \sigma cos(\theta_{eq}) g'(C^{n+1})\right]\frac{\partial C^{n+1}}{\partial x_t} \frac{l_s(\gamma+1)dn }{\mu (l_s+\gamma dn)}.
 \end{gathered}
\end{equation}\end{linenomath}

By transforming back from boundary-fitted ($n, t$) to the cartesian coordinates $(X_1, X_2)$, the updated value of $\tilde{u}^{**}_i$ at the ghost points can be found
\begin{linenomath}\begin{equation} \label{UpdateVandW}
\begin{gathered}
\tilde{u}^{**}_1 = V^{**}_{n_G} n_1+ V^{**}_{t_G} n_2,  \\
\tilde{u}^{**}_2 = V^{**}_{n_G} n_2+ V^{**}_{t_G} n_1. 
 \end{gathered}
\end{equation}\end{linenomath}
Finally, the velocity components at the corresponding faces are found by interpolating the cell-centred values.

\subsection{Correction step}
For the current implementation, we follow the approach in  \cite{DODD2014416} to correct the calculated second prediction velocity and satisfy the divergence free condition when the density is not uniform. First, we update the pressure field by  solving the following equation:
\begin{linenomath}\begin{equation} \label{PressurePoisson}
\begin{gathered}
\frac{\partial^2}{\partial x_i \partial x_i} P^{n+1} = \frac{\partial}{ \partial x_i} \left[(1-\frac{\rho_0}{\rho^{n+1}})  \frac{\partial}{\partial x_i} \hat{P} \right] + \frac{\rho_0}{\Delta t} \frac{\partial}{\partial x_i} {u^{**}_i},
 \end{gathered}
\end{equation}\end{linenomath}
where $\rho_0 = min(\rho_1,\rho_2)$ and $\hat{P} = 2P^n-P^{n-1}$.  The choice of the numerical algorithm for solving the Poisson's equation depends on the problem setup and the implementation. For instance, we employed a fast Fourier transform to solve equation \ref{PressurePoisson} which requires periodicity in the flow direction. However, any other algorithm (such as iterative methods, multi-grid approach, etc,) can be used to solve the pressure equation.

\begin{figure} [H]
\begin{center}
\includegraphics[width=0.8\textwidth]{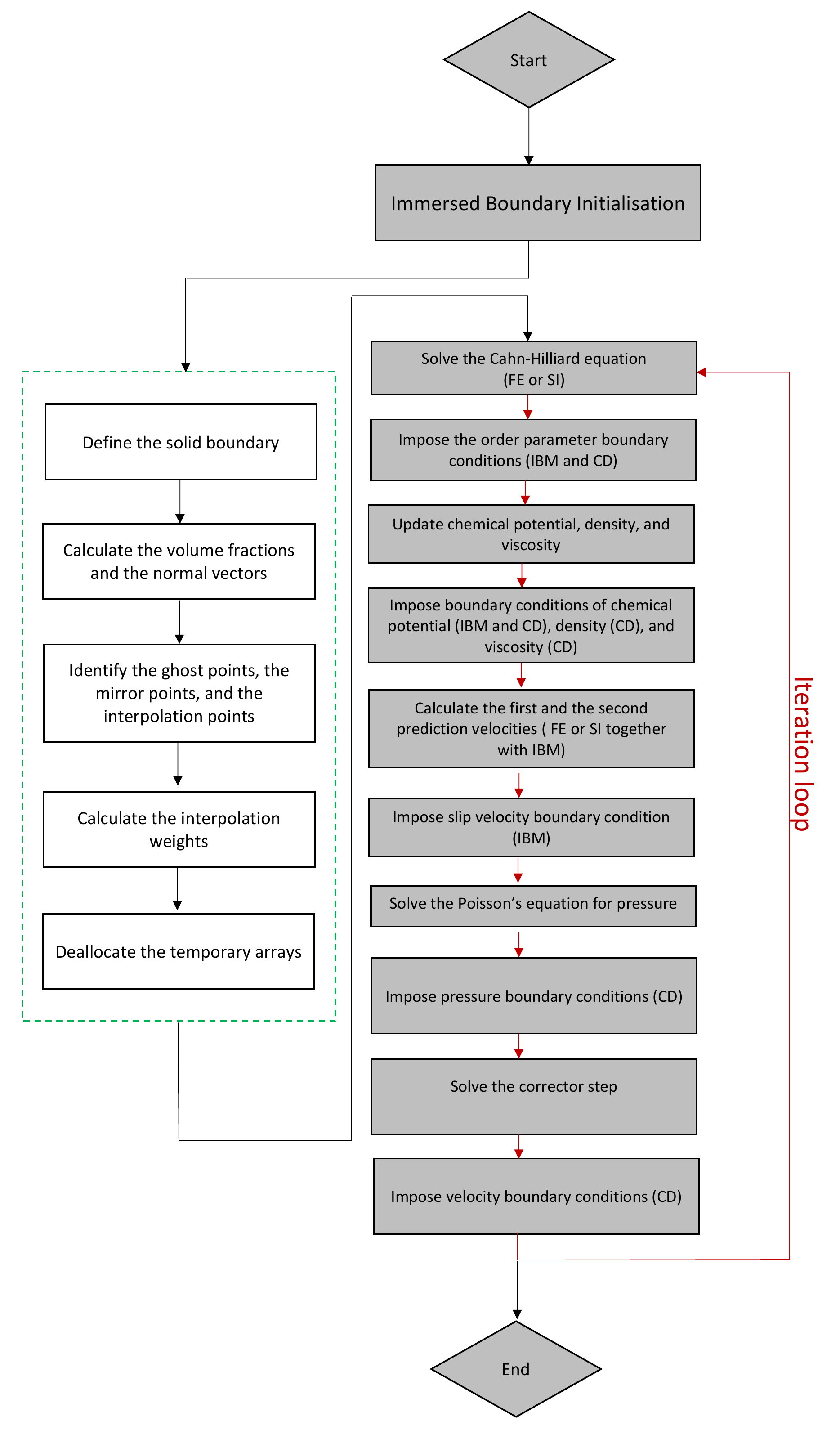}
\put(-325,70){\small FE: Fully Explicit}
\put(-325,50){\small SI: Semi Implicit}
\put(-325,30){\small CD: Computational Domain}
 \end{center}
\caption{Block diagram of the Hybrid IBM-PFM algorithm.}
\label{fig:Chart}
\end{figure}

Having updated the pressure field, we correct the second prediction velocity and calculate the divergence-free velocity as follows:
\begin{linenomath}\begin{equation} \label{correction}
\begin{gathered}
u^{n+1}_i = u^{**}_i - \Delta t \left[\frac{1}{\rho_0} \frac{\partial}{\partial x_i}P^{n+1}  + \left(   \frac{1}{\rho^{n+1}}-  \frac{1}{\rho_0}\right)  \frac{\partial}{\partial x_i}\hat{P} \right].
 \end{gathered}
\end{equation}\end{linenomath}
Details of the algorithm can be found in the above reference.  The algorithm proposed here, with the different steps, is summarised in figure \ref{fig:Chart}.

\section{Numerical tests} \label{Sec:Tests}
To validate the developed code and the numerical model, different simulations are performed. First, we validate the PFM module of the code by comparing results for a droplet spreading on a flat wall and a two phase Couette flow in a channel against previous studies \citep{Nakamura2013,Bao2012}. Next, simulations of the phase separation problem presented in \cite{Nishida2018} and of droplet spreading on an inclined flat wall are performed to validate the hybrid IBM-PFM algorithm. To test the new approach in more complex geometries, 
we simulate a droplet spreading over 2 sinusoidal walls and finally present the results pertaining a three-dimensional droplet spreading over a three-dimensional solid curved wall.

\subsection{Droplet spreading on a flat plate} \label{SpreadingNakaPFM}
\begin{figure} [H]
\centering
\includegraphics[width = 0.65\textwidth]{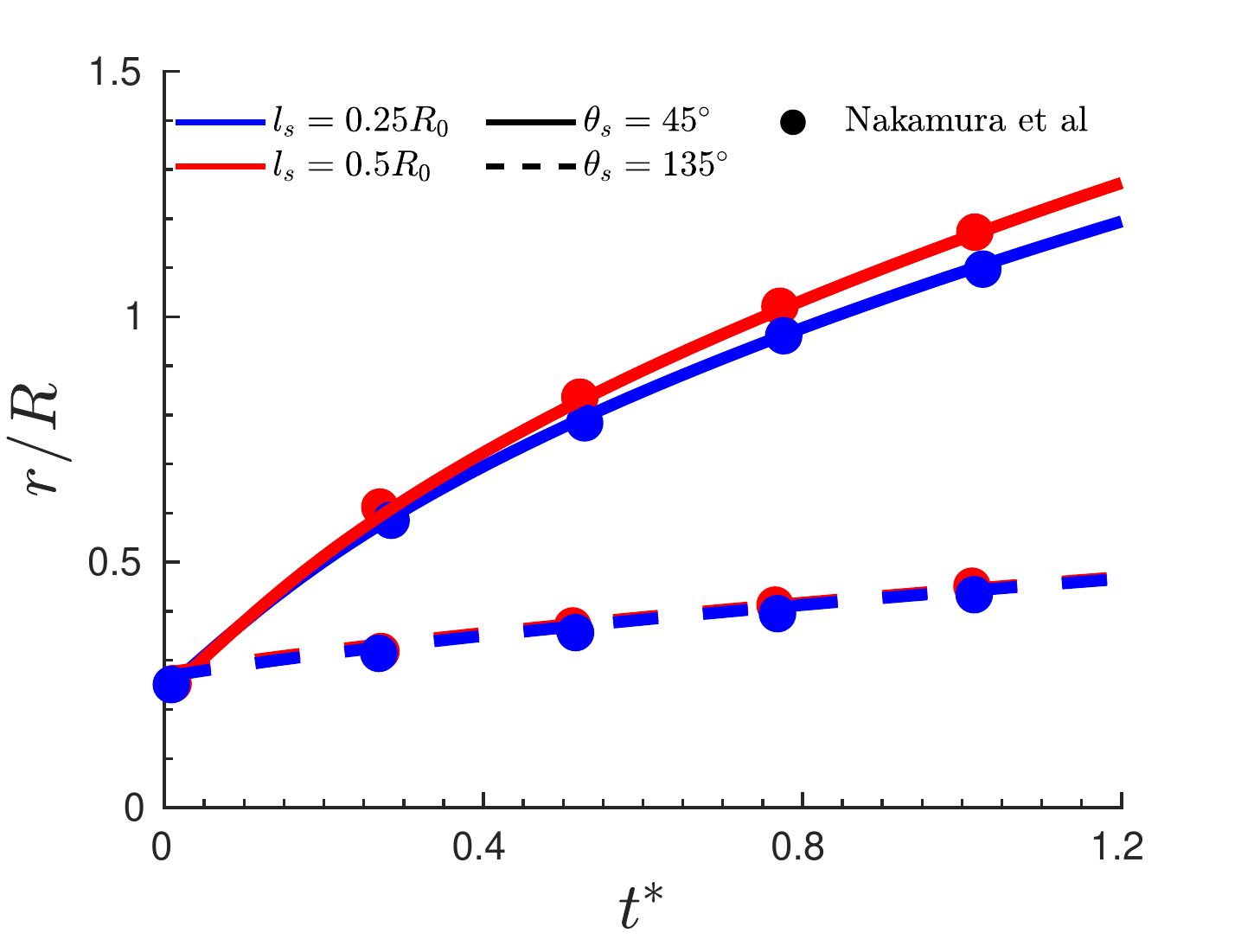}
%\begin{picture}(0,0)
  %\put(-245,111){\includegraphics[height=1.cm]{validations/Droplet/f0.png}}
  %\put(-111,181){\includegraphics[height=1.cm]{validations/Droplet/f45slip1.png}}
 %\end{picture}
\caption{Evolution of the normalized wetting radius ($r/R$) versus the non-dimensional time ($t^*=t/{(\rho R^3/ \sigma)}^{(1/2)}$). The solid lines indicate the results of the present simulations with $\theta_{eq}=45^\circ$, the dashed lines those with $\theta_{eq}=135^\circ$, and the symbols the data in \cite{Nakamura2013} for the same configuration. The blue and the red colours pertain cases with wetting radius $l_s=0.25R$ and $l_s=0.5R$. } 
\label{fig:WettingRadiusPFM}
\end{figure}

A two-dimensional circular droplet is initially placed just above a flat wall so that the interface is almost tangent to the wall.  \cite{Nakamura2013} assumed that the droplet is so small that the gravitational forces are negligible compared to the surface tension forces. The Reynolds ($Re$), capillary ($Ca$), and Cahn numbers ($Cn$) are defined 
with reference to the initial droplet radius ($R$), the reference velocity ($U_{ref} = \sigma/\mu$), the density of the liquid phase ($\rho_L$), the surface tension coefficient ($\sigma$), and the interface thickness ($\epsilon$). The density and viscosity ratio between the two phases are equal to $100$ and
\begin{linenomath}\begin{equation} \label{DropletParameter}
\begin{gathered}
 Re = \frac{\rho U_{ref} R}{\mu} = 4,  \, Ca= \frac{\mu U_{ref} }{\sigma}=1,\,  Cn=\frac{\epsilon}{R}=0.022.
 \end{gathered}
\end{equation}\end{linenomath}
Note that throughout the paper we use the same definition for all the non-dimensional parameters as in equation \ref{DropletParameter}.  In this simulation, the wall friction coefficient $\mu_f$ is set to zero. The computational domain is $[0,10R] \times [0,10R]$ and the number of grid  points per droplet diameter is equal to 320. Figure  \ref{fig:WettingRadiusPFM} shows the evolution of the normalised wetting radius versus the non-dimensional time ($t^*=t/{(\rho R^3/ \sigma)}^{(1/2)}$) for two different contact angles ($\theta_{eq}=45^\circ
$ and $135^\circ$) and two different slip lengths ($l_s=0.25R$ and $0.5R$). The solid lines represent the results of the present simulations with $\theta_{eq}=45^\circ$, the dashed lines those with $\theta_{eq}=135^\circ$, and the symbols the results in \cite{Nakamura2013}. The blue and the red colours illustrate different wetting radii, $l_s=0.25R$ and $l_s=0.5R$. This simulation used the explicit time integration with time step equal to $10^{-5}$.
The proper choice of the time step depends on the three physical time scales of the problem, namely, the convective time scale ($\Delta t_C$), the viscous time scale ($\Delta t_{\nu}$), and surface tensions time scale ($\Delta t_{\sigma}$) and a safety factor which is required for the stability of the Cahn-Hilliard equation ($\approx 0.1$). We. estimate the time step of the simulations as follows \citep{DODD2014416}:
\begin{linenomath}\begin{equation} \label{timestep}
\begin{gathered}
 \Delta t_c = \frac{\Delta X}{{|U|}_{max}}, \\
 \Delta t_\nu = \frac{Re {\Delta X}^2}{6}, \\
 \Delta t_\sigma = \sqrt{\frac{Re  Ca (\rho_1+\rho_2) {\Delta X}^3}{4\pi}}, \\
 \Delta t \le \frac{1}{10} min(\Delta t_c, \Delta t_\nu,\Delta t_\sigma )
 \end{gathered}
\end{equation}\end{linenomath}
 As shown in the figure, the results of our simulations are in good agreement with those by \cite{Nakamura2013}.

\subsection{Two phase Couette flow}
As the second validation for the PFM module, a two phase Couette flow is simulated and the results compared with those by \cite{Bao2012}. The simulation domain is $[0,1] \times [0,0.25]$. Two interfaces are initially located at $x=0.25$ and $x=0.75$ with a tangent hyperbolic variation of the order parameter from one phase to the other according to:
\begin{linenomath}\begin{equation} \label{InitializingC}
\begin{gathered}
 C(X_1,X_2,t=0) = \tanh \left( \frac{1}{\sqrt 2 Cn}(0.25 L_{X_1}- |X_1-0.5 L_{X_1}|) \right).
 \end{gathered}
\end{equation}\end{linenomath}
In this simulation, the wall velocity is $u_w=0.2$, $Re=5$ (defined as above with $U_{Ref}=u_w$), $l_s=0.0025$, $Cn=0.004$, and $ Ca=1/12$. As in \cite{Bao2012}, the mobility coefficient is defined in non-dimensional form $M^*= (3M\sigma)/(2\sqrt(2)U_{ref} {L_{ref}}^2)=0.0005$ and the wall friction coefficient is expressed in terms of the relaxation time $V_s= (3\sigma \Gamma_r L_{ref})/(2\sqrt(2) U_{ref})=200$, where $L_{ref}$ is the channel height and $\Gamma = 1/(\mu_f \epsilon)$).  We solved the system of equations using the explicit approach and with a time step equal to $10^{-5}$.

\begin{figure}[H]
\centering
\includegraphics[width = 0.63\textwidth]{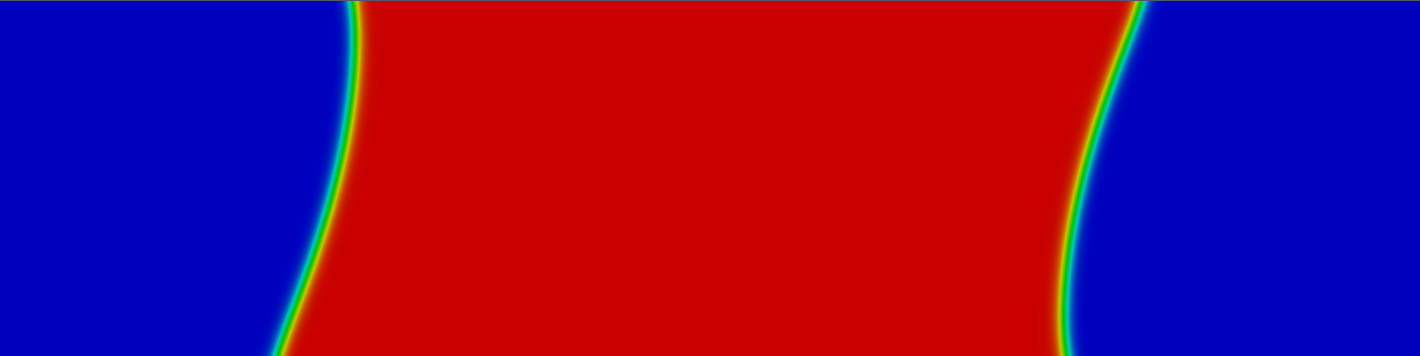}
\includegraphics[width = 0.7\textwidth]{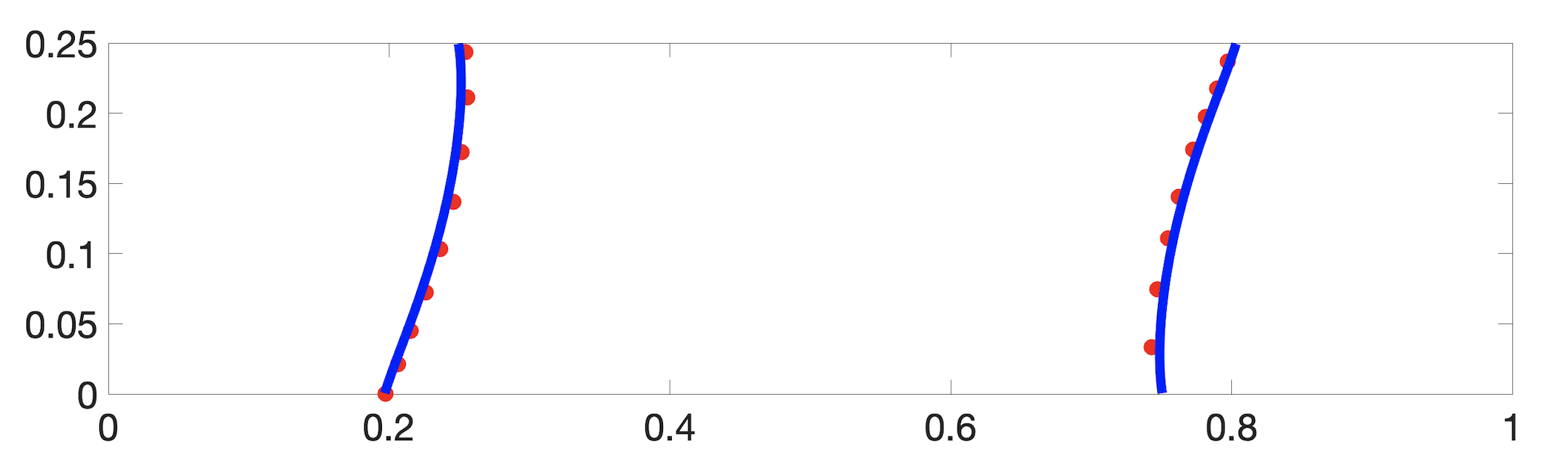}
\caption{top: Contour of the order parameter at $t^*=1.875$. Bottom: interface profile at $t^*=1.875$. The blue lines represents the results of our simulation and the red dots those of \cite{Bao2012}. }
\label{CouetteContour}
\end{figure}

We report in figure \ref{CouetteContour} the contour of the order parameter at $t^*=1.875$ and, for an easier comparison, the digitised interface location from the simulation by \cite{Bao2012} in a second panel. The blue lines represent the results of our simulation and the red dots those by \cite{Bao2012}, with a good match between the two data sets. 

\subsection{Phase separation problem}
To validate the complete hybrid IBM-PFM model, first, we consider a phase separation problem in a square box. The computational domain has size $[0,1] \times [0,1]$ and the phase field order parameter is initialised in the same way as in the simulations by \cite{Nishida2018}:
\begin{linenomath}\begin{equation} \label{InitializingCphaseSep}
\begin{gathered}
 C(X_1,X_2,t=0) = 0.5+0.12\cos(2 \pi X_1)\cos(2 \pi X_2)+ 0.2\cos(\pi X_1)\cos(3\pi X_2)
 \end{gathered}
\end{equation}\end{linenomath}
Two cases are simulated: in the first case, the phase separation occurs in the absence of any velocity field, while in the second one, the initial velocity field is initialised as
\begin{linenomath}\begin{equation} \label{InitializingVphaseSep}
\begin{gathered}
 u_1(X_1,X_2,t=0)= -{\sin}^2(\pi X_1)\sin(2\pi X_2), \\
 u_2(X_1,X_2,t=0)= {\sin}^2(\pi X_2)\sin(2\pi X_1).
 \end{gathered}
\end{equation}\end{linenomath}
\begin{figure}[H]%[H]
\centering

\includegraphics[width = 0.18\textwidth]{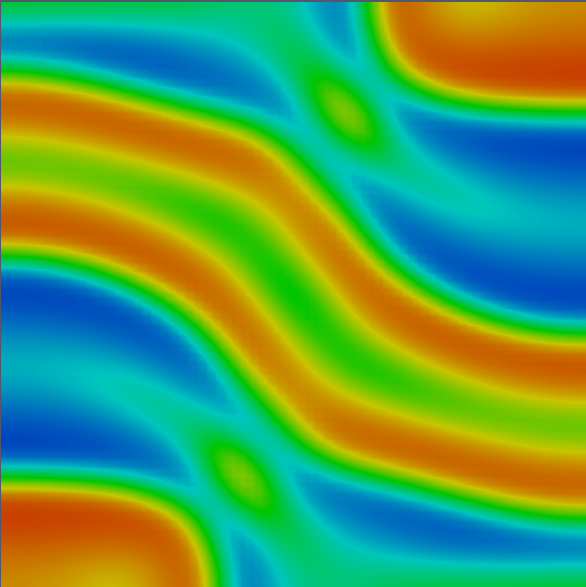}
\includegraphics[width = 0.18\textwidth]{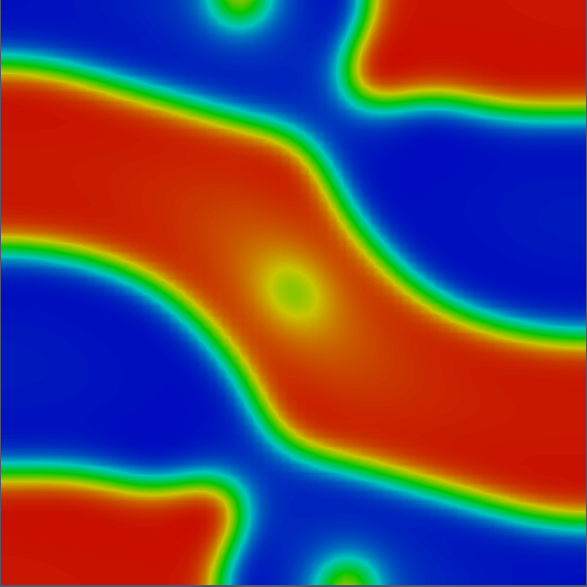}
\includegraphics[width = 0.18\textwidth]{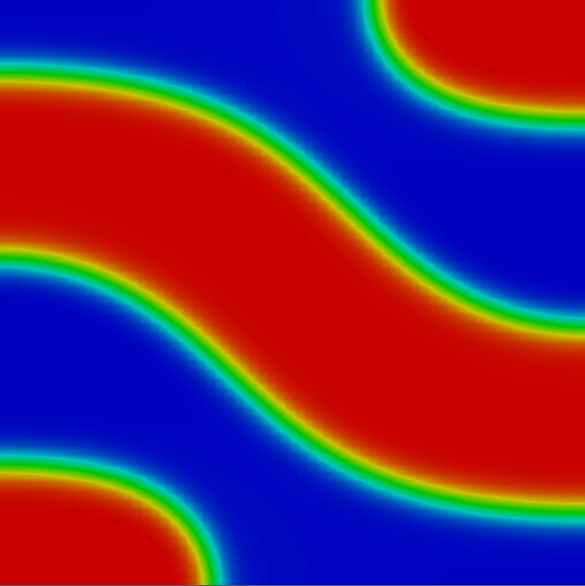}
\includegraphics[width = 0.18\textwidth]{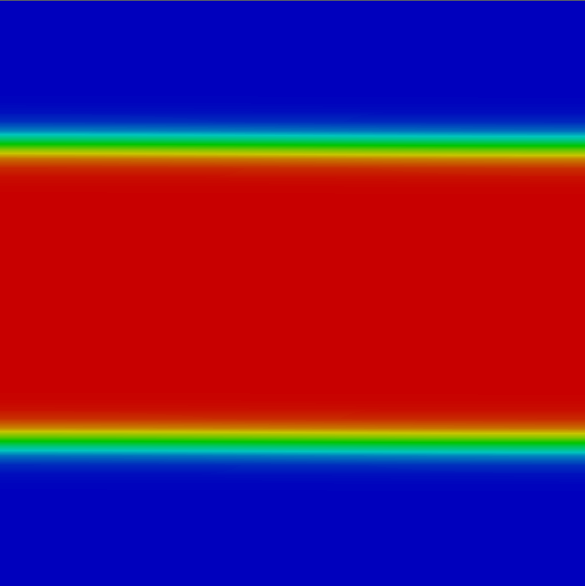}

\includegraphics[width = 0.18\textwidth]{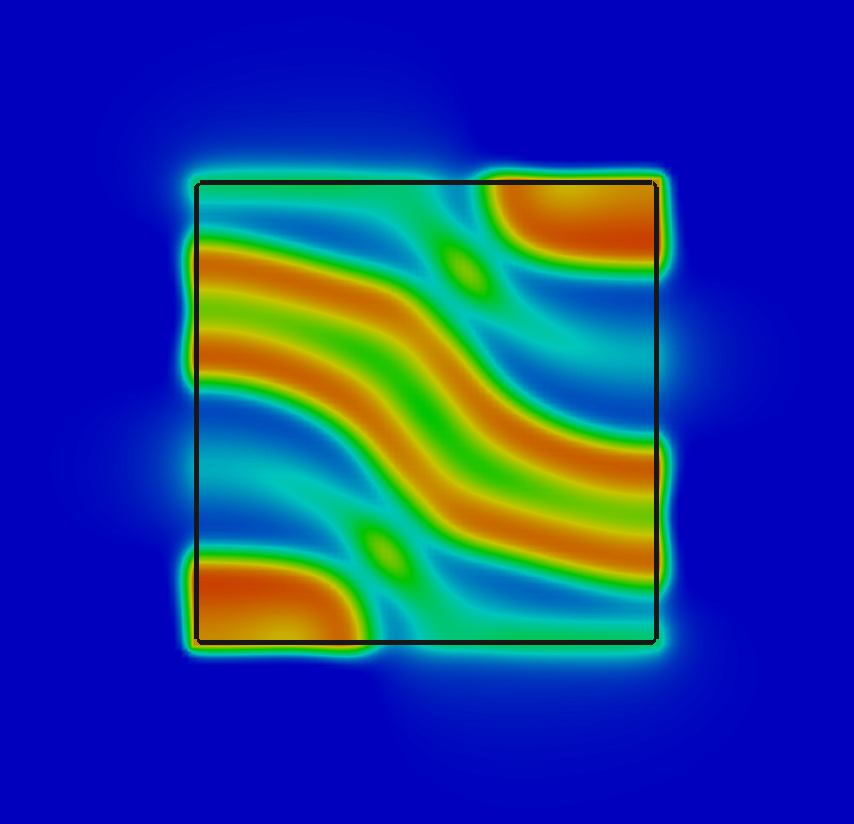}
\includegraphics[width = 0.18\textwidth]{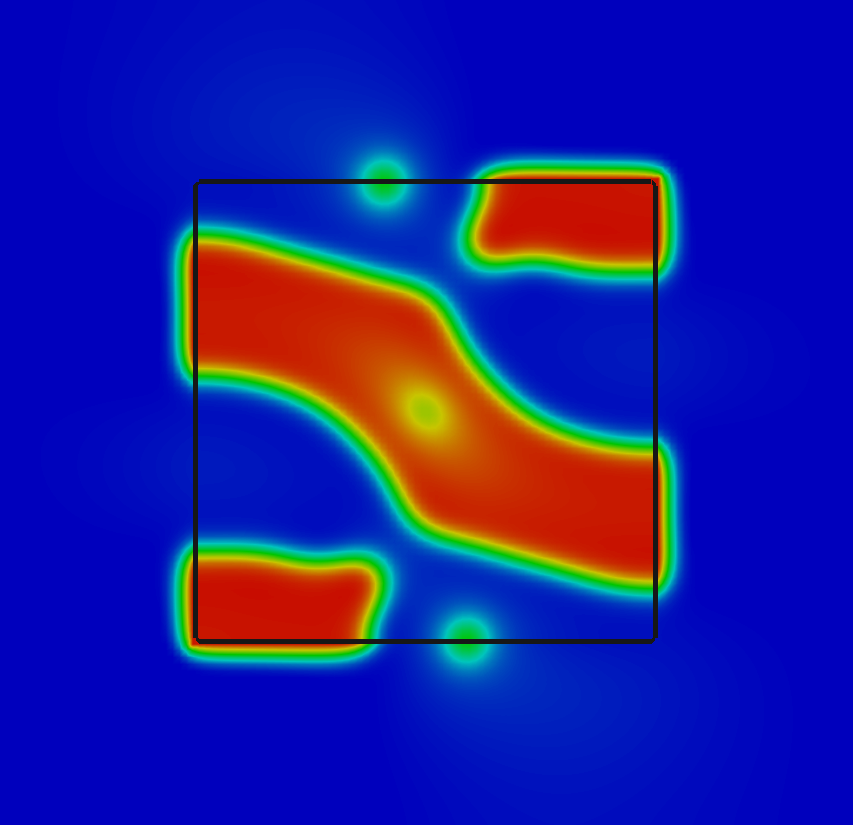}
\includegraphics[width = 0.18\textwidth]{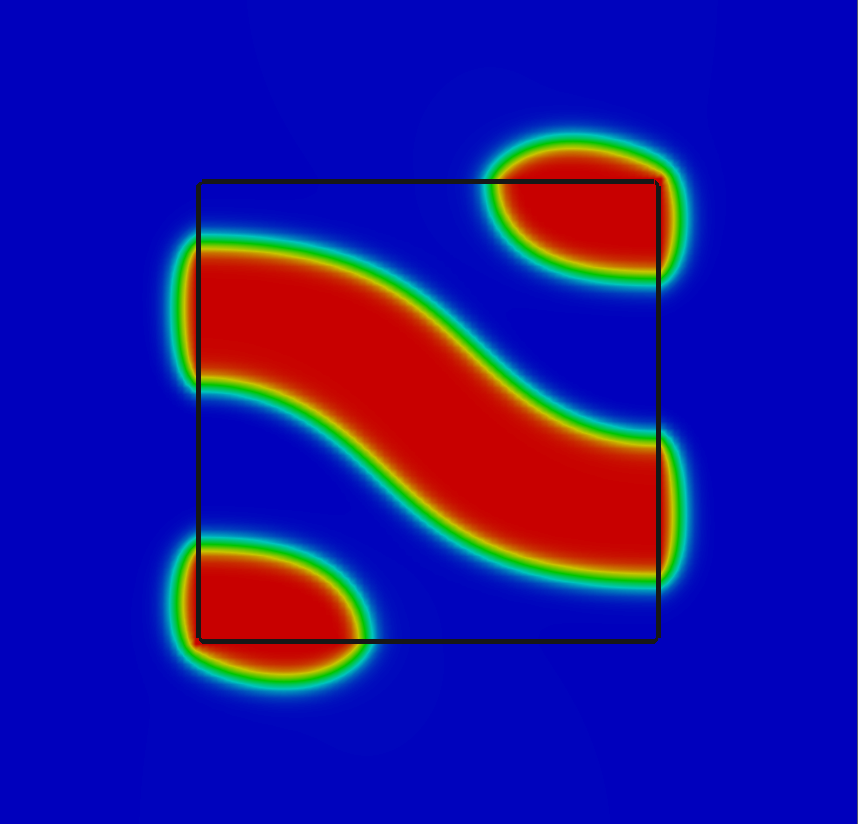}
\includegraphics[width = 0.18\textwidth]{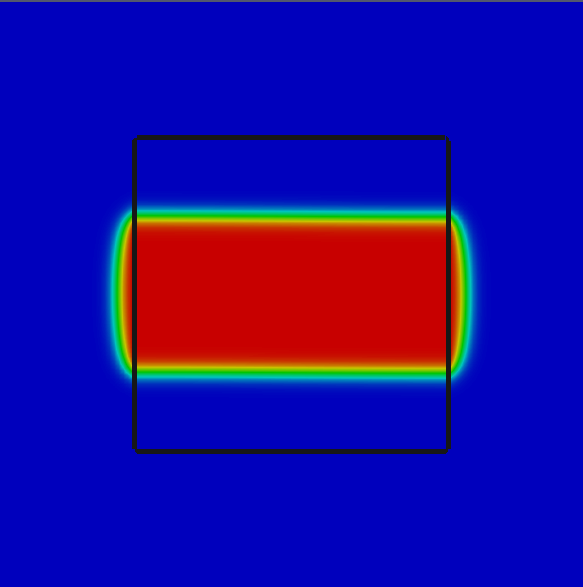}

\includegraphics[width =  0.18\textwidth]{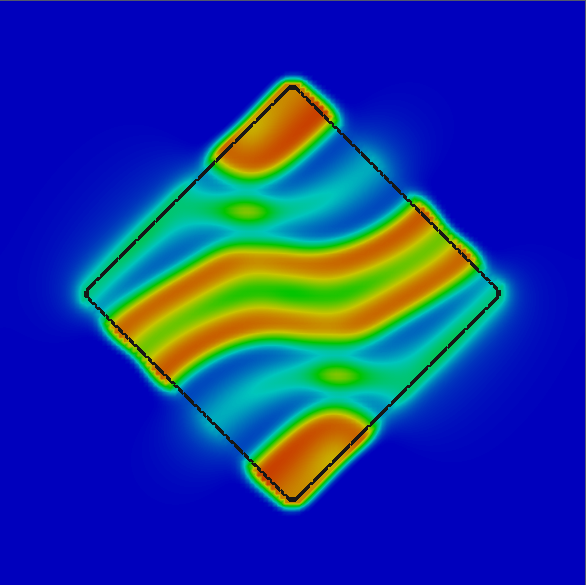}
\includegraphics[width =  0.18\textwidth]{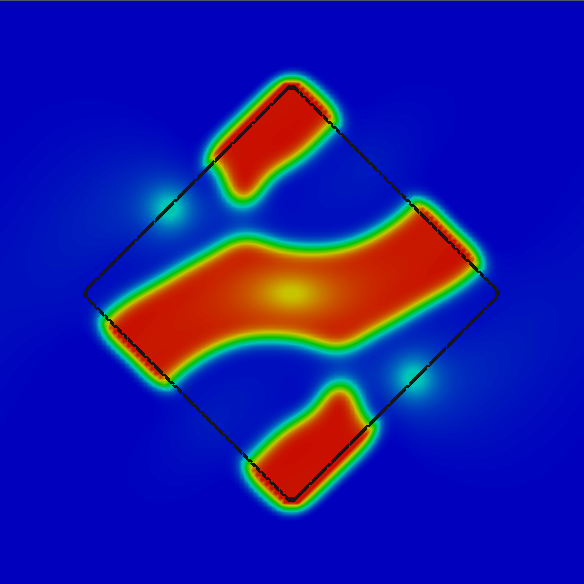}
\includegraphics[width =  0.18\textwidth]{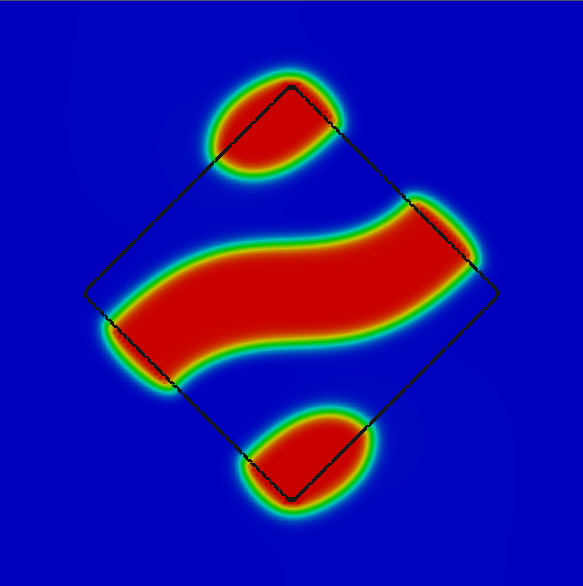}
\includegraphics[width =  0.18\textwidth]{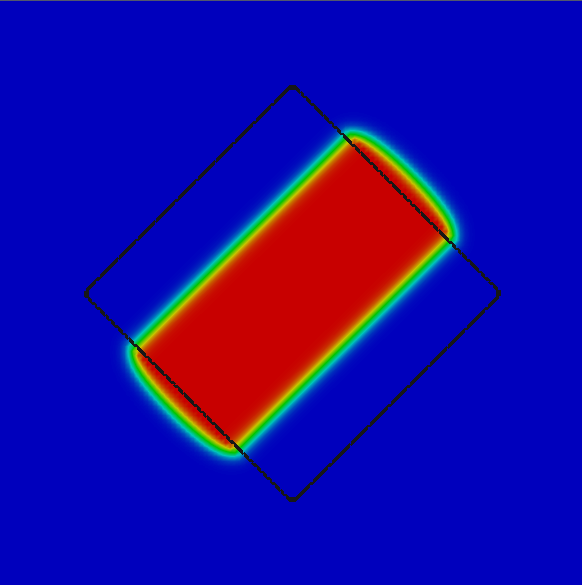}
\put(-250,-12){\small $t^*=0.04$}
\put(-181,-12){\small $t^*=0.1$}
\put(-114,-12){\small $t^*=1.0$}
\put(-47,-12){\small $t^*=45$}
\caption{Contours of the order parameter for the phase separation in the absence of flow and with contact angle equal to $90^\circ$. The first row represents the results of the PFM code, the second and the third rows are the results of the PFM-IBM code with a box rotation angle equal to $\alpha_r= 0^\circ$ and $45^\circ$, where the thin solid line indicates the location of the walls modelled by the IBM. The non-dimensional times in the first, second, third, and fourth column are $t^* = 0.04, 0.1, 1.0$, and $45.0$. }
\label{ContourNoflow}
\end{figure}
No-slip velocity boundary conditions are imposed on the four sides of the simulation box. The Reynolds and capillary numbers are set equal to $100$ and $1$. 
The mobility coefficient ($M$) is defined using a Peclet number, $Pe=(2\sqrt 2 U_{ref}\epsilon L_{ref})/(3 M \sigma)=1$ where $U_{ref}=\sigma/\mu$ and $L_ref$ is the box height. 
The contact angle, wall friction coefficient, and slip length are set equal to $\theta_{eq}=90^\circ$, $\mu_f=0$, and $l_s=0$. 
For each of the 2 phase separation problems (with and without flow), the simulations are first performed with the previously validated PFM code (without any immersed boundary), using the explicit time integration with time step equal to $10^{-5}$. Then, we define the square box boundaries by means of the IBM with two different rotation angles, $\alpha_r=0^\circ$ and $45^\circ$. For the IBM simulations, the time step is reduced to $10^{-6}$. Figure \ref{ContourNoflow} shows the contours of the order parameter extracted from our PFM simulations and from the two PFM-IBM simulations for the case without flow at four different non-dimensional times. 
The results of the hybrid PFM-IBM code are in good agreement with those of the validated PFM code with wall conditions imposed directly at the domain boundaries.

The total energy of the system is calculated and compared with the one obtained by \cite{Nishida2018}. This  is defined as the summation of the kinetic and interfacial energy 
\begin{linenomath}\begin{equation} \label{TotalEnergy}
\begin{gathered}
 E(t)) = \int_{\Omega} \frac{1}{2} \left( {u_i}{u_i}\right)dV + \frac{1}{Re \, Ca \,\epsilon}\int_{\Omega} \left(\frac{1}{4}C^2{(1-C)}^2 + \frac{\epsilon^2}{2} \frac{\partial C}{\partial x_i} \frac{\partial C}{\partial x_i}\right)dV.
 \end{gathered}
\end{equation}\end{linenomath}

\begin{figure} [H]
\centering
\includegraphics[width = 0.85\textwidth]{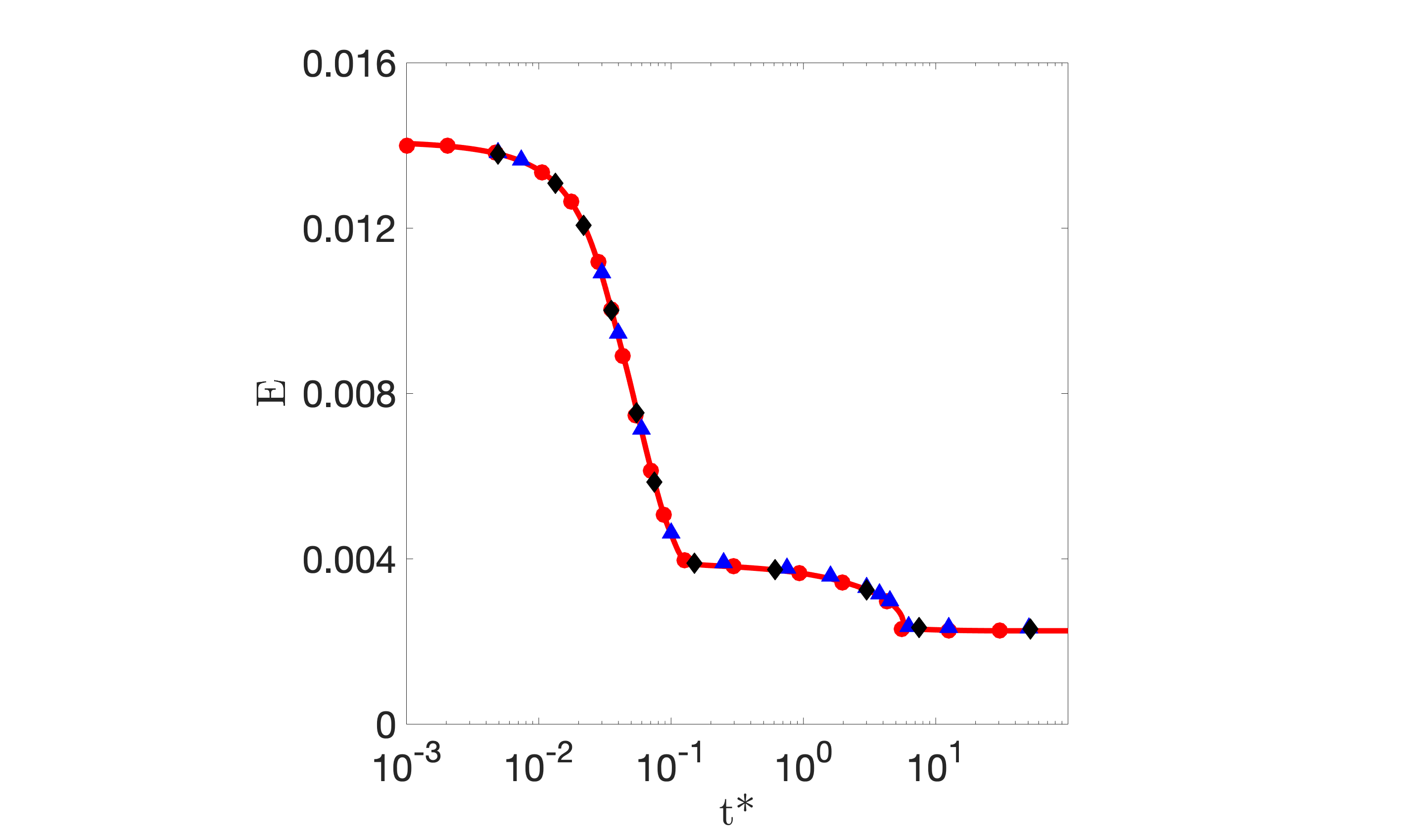}
\caption{Evolution of the total energy of the system for the phase separation problem in the absence of flow. Red solid line indicates the results of our PFM code, the red dots are the results by \cite{Nishida2018}, and the blue and black symbols the results of our hybrid PFM-IBM code for $\alpha_r=0^\circ,$ and $45^\circ$, respectively. }
\label{TotalEnergynoFlow}
\end{figure}

The evolution of the total energy is depicted in figure \ref{TotalEnergynoFlow} for the different simulations: the red solid line shows the results of our PFM simulations, the red dots are the results \cite{Nishida2018}, and the blue and black symbols represent the results of our hybrid PFM-IBM code for box rotations $\alpha_r=0^\circ,$ and $45^\circ$.

\begin{figure} [H]
\centering
\includegraphics[width = 0.18\textwidth]{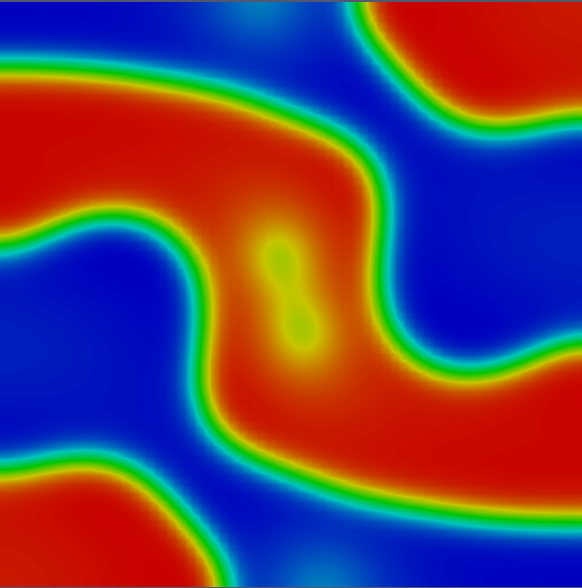}
\includegraphics[width = 0.18\textwidth]{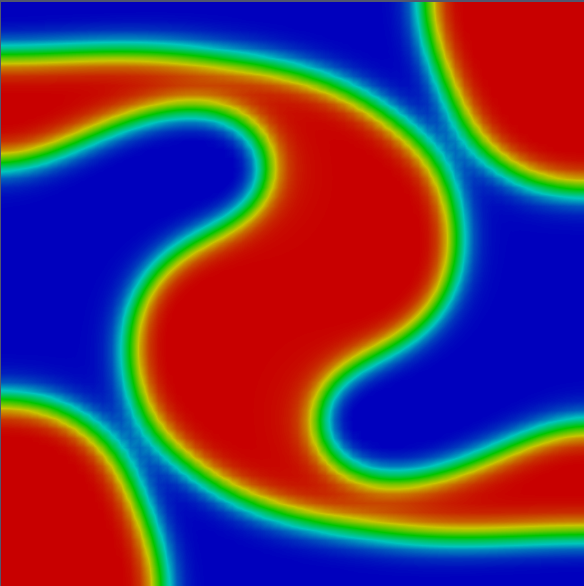}
\includegraphics[width = 0.18\textwidth]{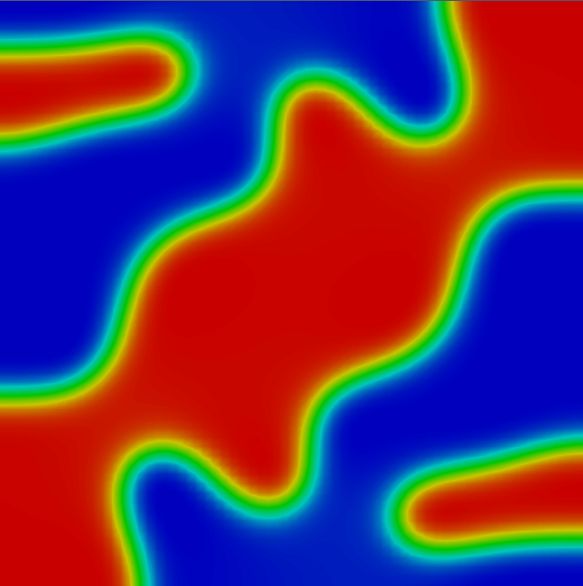}
\includegraphics[width = 0.18\textwidth]{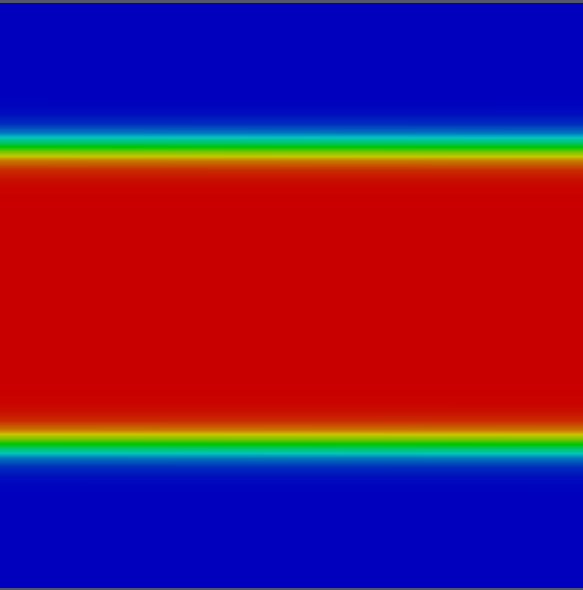}

\includegraphics[width = 0.18\textwidth]{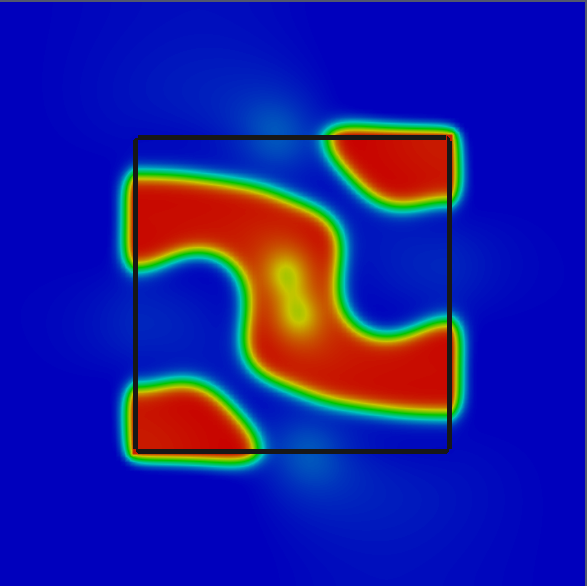}
\includegraphics[width = 0.18\textwidth]{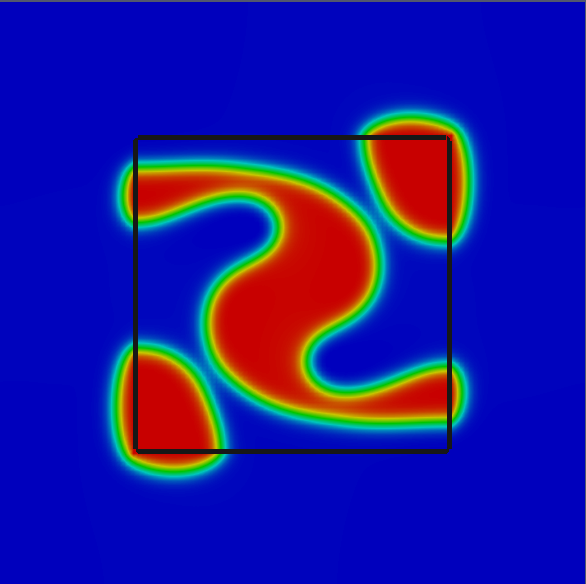}
\includegraphics[width = 0.18\textwidth]{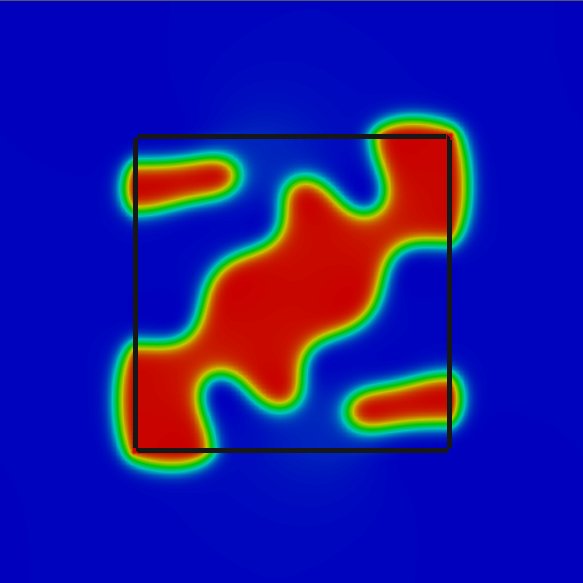}
\includegraphics[width = 0.18\textwidth]{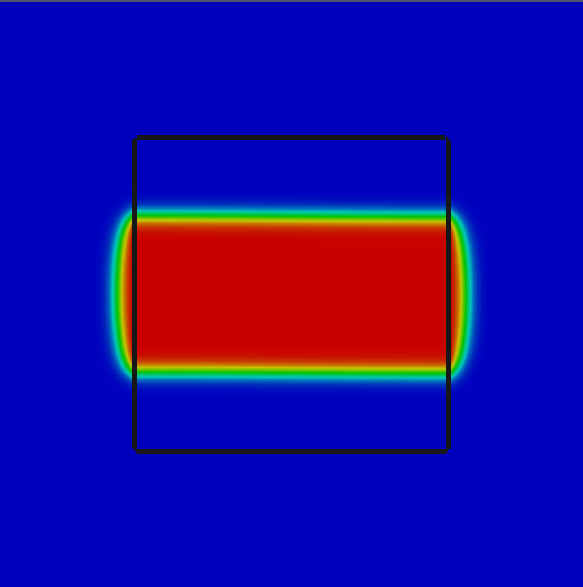}

\includegraphics[width =  0.18\textwidth]{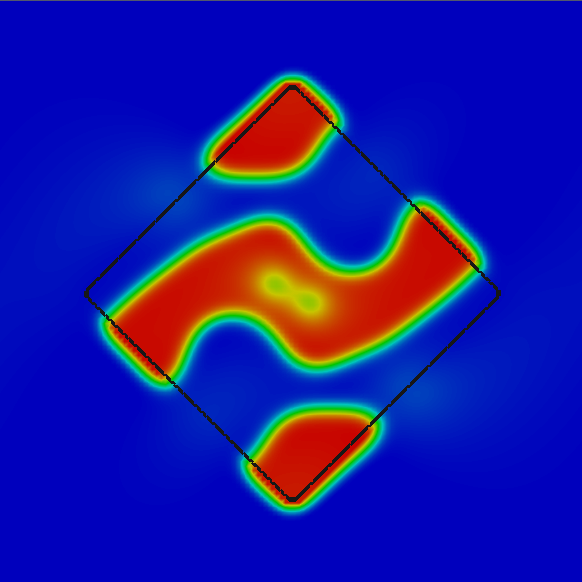}
\includegraphics[width =  0.18\textwidth]{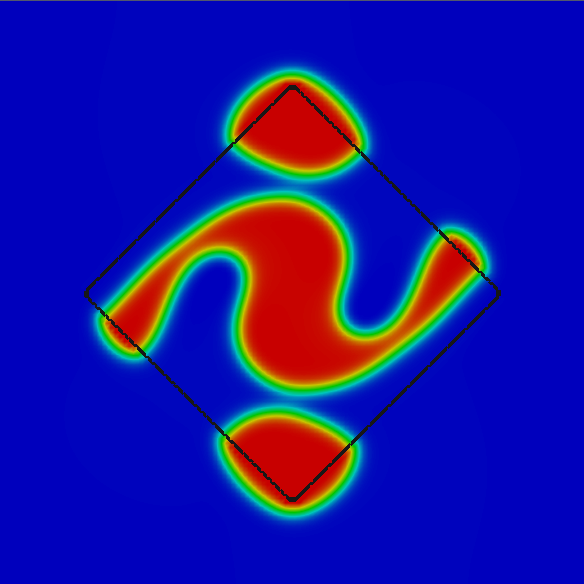}
\includegraphics[width =  0.18\textwidth]{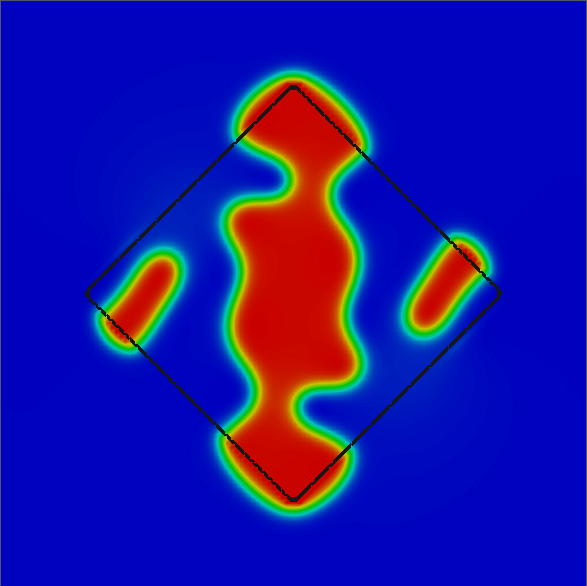}
\includegraphics[width =  0.18\textwidth]{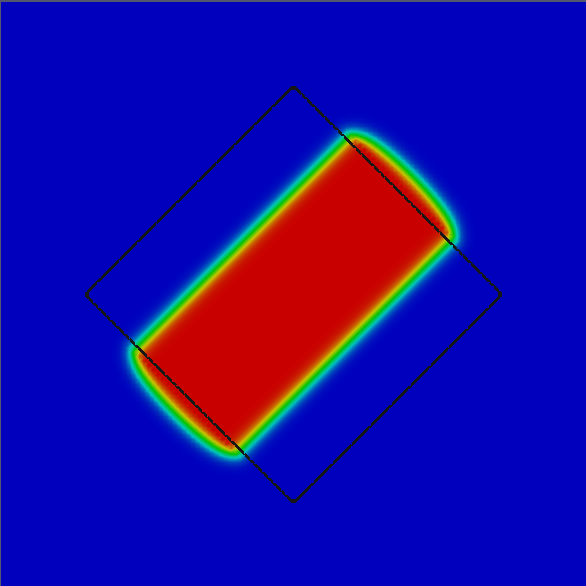}
\put(-250,-12){\small $t^*=0.1$}
\put(-181,-12){\small $t^*=0.6$}
\put(-114,-12){\small $t^*=0.8$}
\put(-47,-12){\small $t^*=45$}
\caption{Contours of the order parameter for two-fluid phase separation with flow and  contact angle equal to $90^\circ$. The first row represents the results of the PFM code, the second and the third  rows are the results of the PFM-IBM code with a box rotation angle of $\alpha_r= 0^\circ$ and $45^\circ$, where the thin solid line indicates the location of the walls modelled by the IBM. The non-dimensional times in the first, second, third, and fourth column are $t^* = 0.1, 0.6, 0.8, 45.0$ }
\label{ContourFlow}
\end{figure}

\begin{figure} [H]
\centering
\includegraphics[width = 0.85\textwidth]{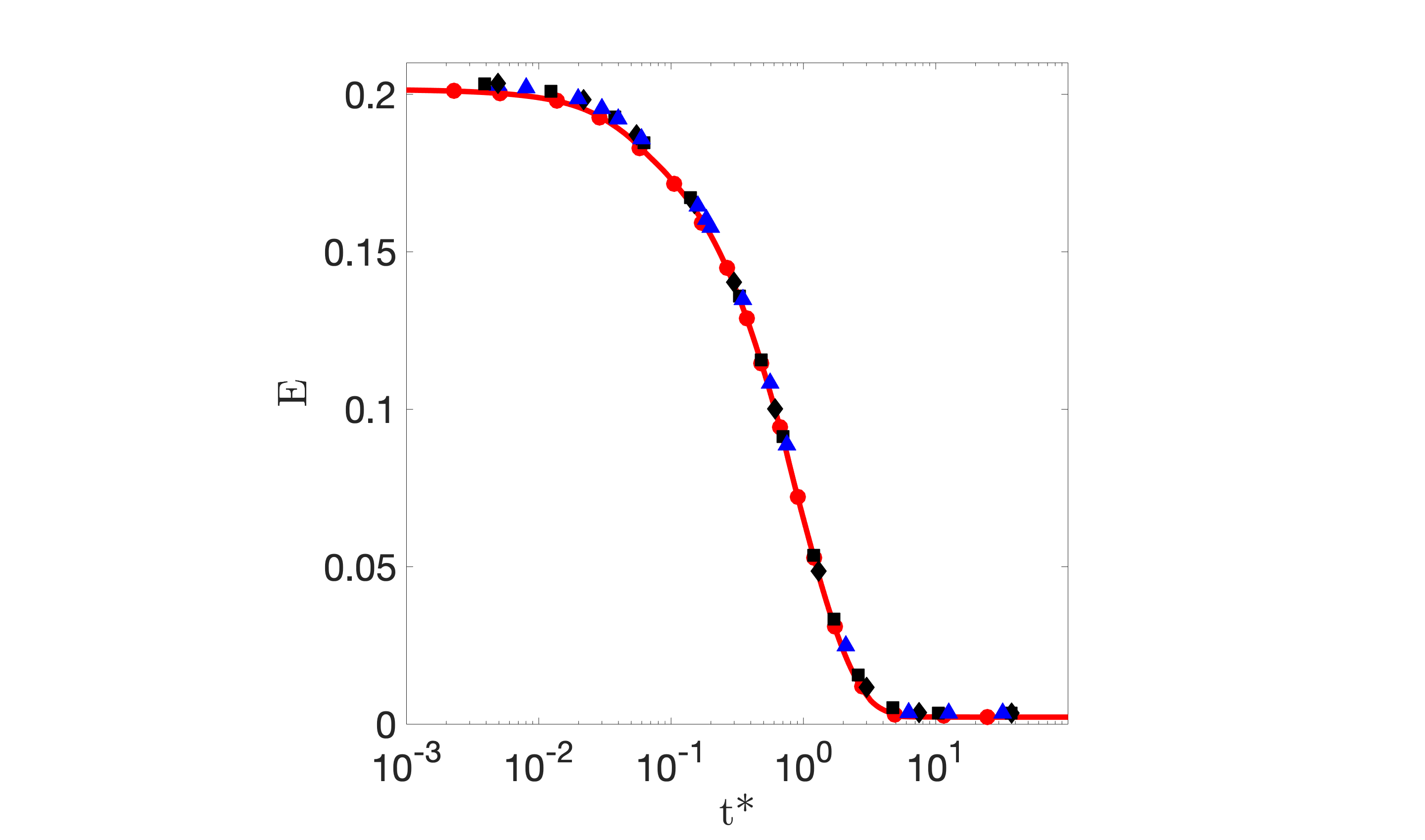}
\caption{Evolution of the total energy of the system for the phase separation problem with flow. The red solid line indicates the results of our PFM code, the red dots are the results by \cite{Nishida2018}. Blue and black symbols represent the results of our hybrid PFM-IBM code for rotations of the computational box $\alpha_r=0^\circ,$ and $45^\circ$. }
\label{TotalEnergyFlow}
\end{figure}

The contours of the order parameter for the phase separation problem with flow are reported in figure \ref{ContourFlow}. Again, the contact angle, wall friction coefficient, and slip length are set equal to $\theta_{eq}=90^\circ$, $\mu_f=0$ , and $l_s=0$.
The corresponding 
evolution of the total energy of the system is presented in figure \ref{TotalEnergyFlow}. According to figure \ref{ContourFlow} and figure \ref{TotalEnergyFlow} good agreements between the results of our PFM code, hybrid PFM-IBM code and those obtained by \cite{Nishida2018} is again achieved.

\subsection{Droplet spreading on a rotated wall without gravity}
\begin{figure} [H]
\centering
\includegraphics[width = 0.9\textwidth]{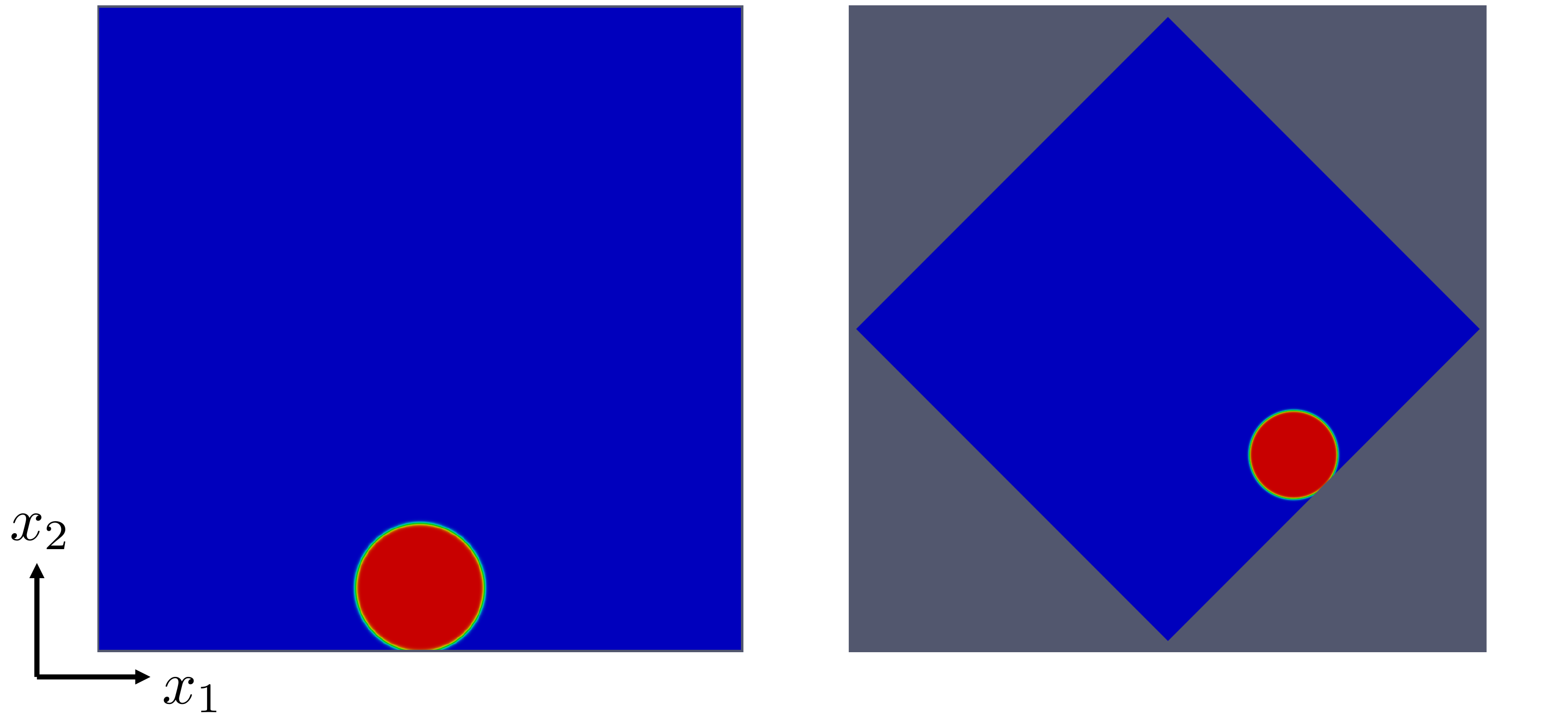}
\caption{Initial configuration of the simulation of a droplet spreading on a flat wall in a rotated computational domain. Left: PFM setup, Right: IBM setup. }
\label{fig:DropletRot}
\end{figure}

To validate our hybrid code for different slip lengths, we reproduce once more the results of a droplet spreading on a flat wall by \cite{Nakamura2013} but modelling the wall with the IBM module and adding a rotation of the computational box  of $\alpha_r=45^\circ$, as shown in figure \ref{fig:DropletRot}. The computational domain is $[0,20R] \times [0,20R]$ and the edge length of the rotated box is $10R$. All the other simulation parameters are the same as previously reported in section \ref{SpreadingNakaPFM}. In this case, both the explicit and the semi-implicit algorithms with time steps equal to $10^{-5}$ and $10^{-4}$, are tested.

\begin{figure} [H]
\centering
\includegraphics[width = 0.65\textwidth]{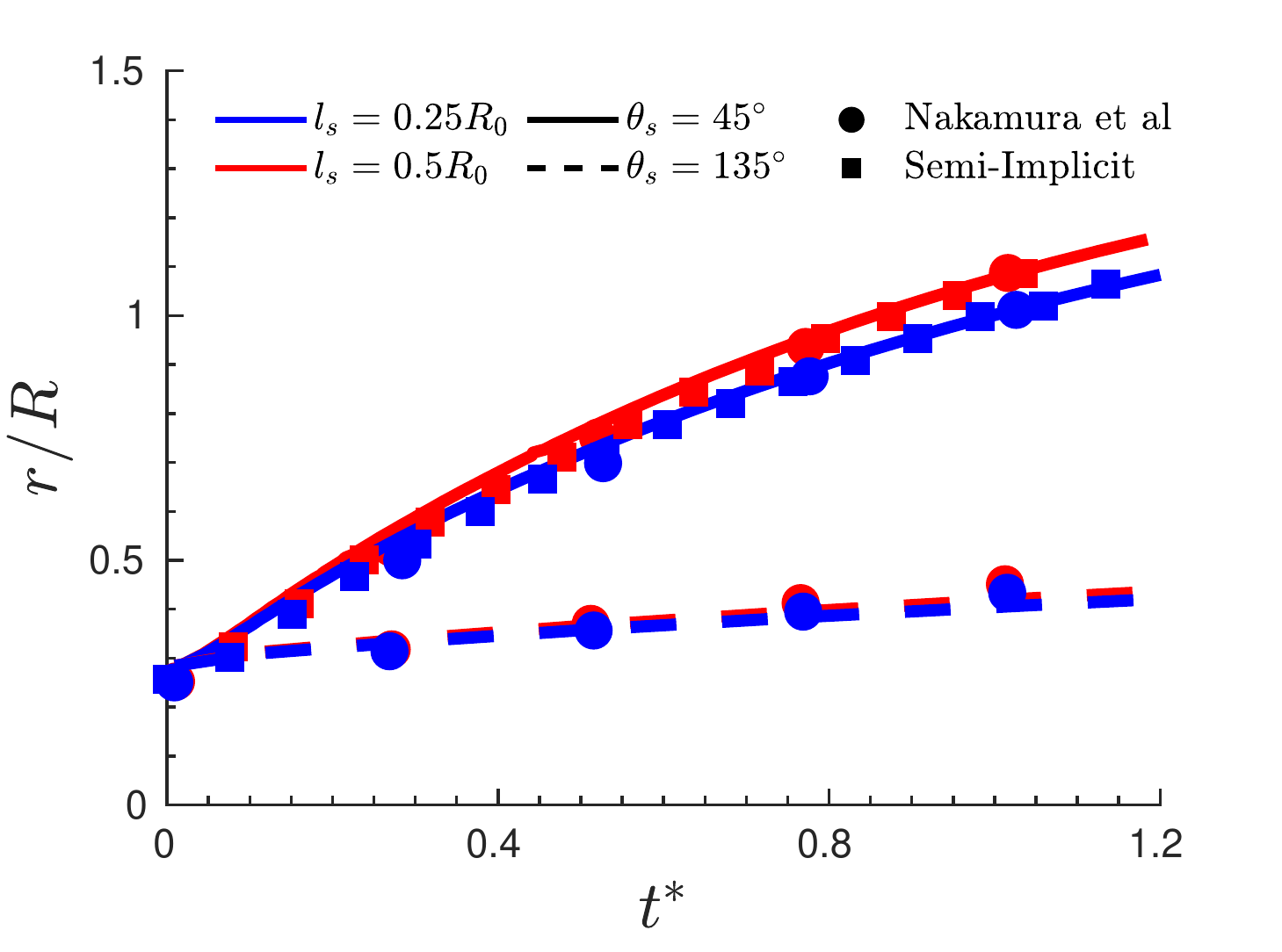}
  %\put(-111,181){\includegraphics[height=1.cm]{validations/Droplet/f45slip1.png}}
\caption{Evolution of the normalised wetting radius ($r/R$) versus the non-dimensional time ($t^*=t/{(\rho R^3/ \sigma)}^{(1/2)}$). Solid lines show the results for $\theta_{eq}=45^\circ$, dashed lines those for  $\theta_{eq}=135^\circ$, the filled circles represent the results by \cite{Nakamura2013}, and the filled squares indicate the results of our simulations using the semi-implicit algorithm. Blue and red colours represent slip lengths $l_s=0.25R$ and $l_s=0.5R$.}
\label{fig:WettingRadius}
\end{figure}

Figure \ref{fig:WettingRadius} displays the evolution of the normalised wetting radius versus the non-dimensional time. Solid lines show the results for $\theta_{eq}=45^\circ$, dashed lines those for $\theta_{eq}=135^\circ$, filled circles the results by \cite{Nakamura2013} and filled squares  our simulations using the semi-implicit algorithm. Blue and red colours represent cases with slip length $l_s=0.25R$ and $l_s=0.5R$, respectively. 

As shown in figure \ref{fig:WettingRadius}, the results of the hybrid PFM-IBM code are in good agreement with the reference data.

\subsection{Droplet spreading over sinusoidal surfaces}
To examine the capability of the developed model to simulate more complex geometries, we simulate a droplet spreading over two sinusoidal surfaces with same wave length but a phase shift of $\pi/2$ with respect to the initial droplet position. The wall geometry is defined by
\begin{linenomath}\begin{equation} \label{SinusoidalWall}
\begin{gathered}
 y(x) = h_w \Delta y \sin\left(\frac{(2n-1)\pi}{L_x}-m\frac{\pi}{2}\right)x+h_b \Delta y,
 \end{gathered}
\end{equation}\end{linenomath}
where $h_w$ is the number of grid points defining the amplitude of the wave, $h_b$ the number of grid point below the centreline of the sinusoidal wave, $n$ the number of peaks of the sinusoidal function and $m=0$ or 1 to shift the wave. For the results presented here, the computational domain  is $[0,20R] \times [0,10R] $, $R$ being the droplet radius, and the Cartesian grid consists of $1280\times640 $ grid points. $h_w$ and $h_b$ are set to 30 and 60, respectively. The non-dimensional parameters pertaining these simulations are $Re =4.0$,  $Ca=1.0$, $Cn=0.02$, and $Pe=2\sqrt{2} U_{ref} R \epsilon/3 M \sigma=10^4$ . Two equilibrium contact angles are considered to model a hydrophobic ($\theta_{eq}=135^\circ$) and a hydrophilic ($\theta_{eq}=45^\circ$) wall. The droplet is initially placed at $(10R,5R+(h_w+h_b)dz)$. The wall friction coefficient and the slip length are zero and a quarter of the droplet radius, $\mu_f=0,l_s=0.25R_0$. The simulations are performed using the explicit code with a time step equal to $10^{-5}$.

\begin{figure} [H]
\centering
\includegraphics[width = 0.85\textwidth]{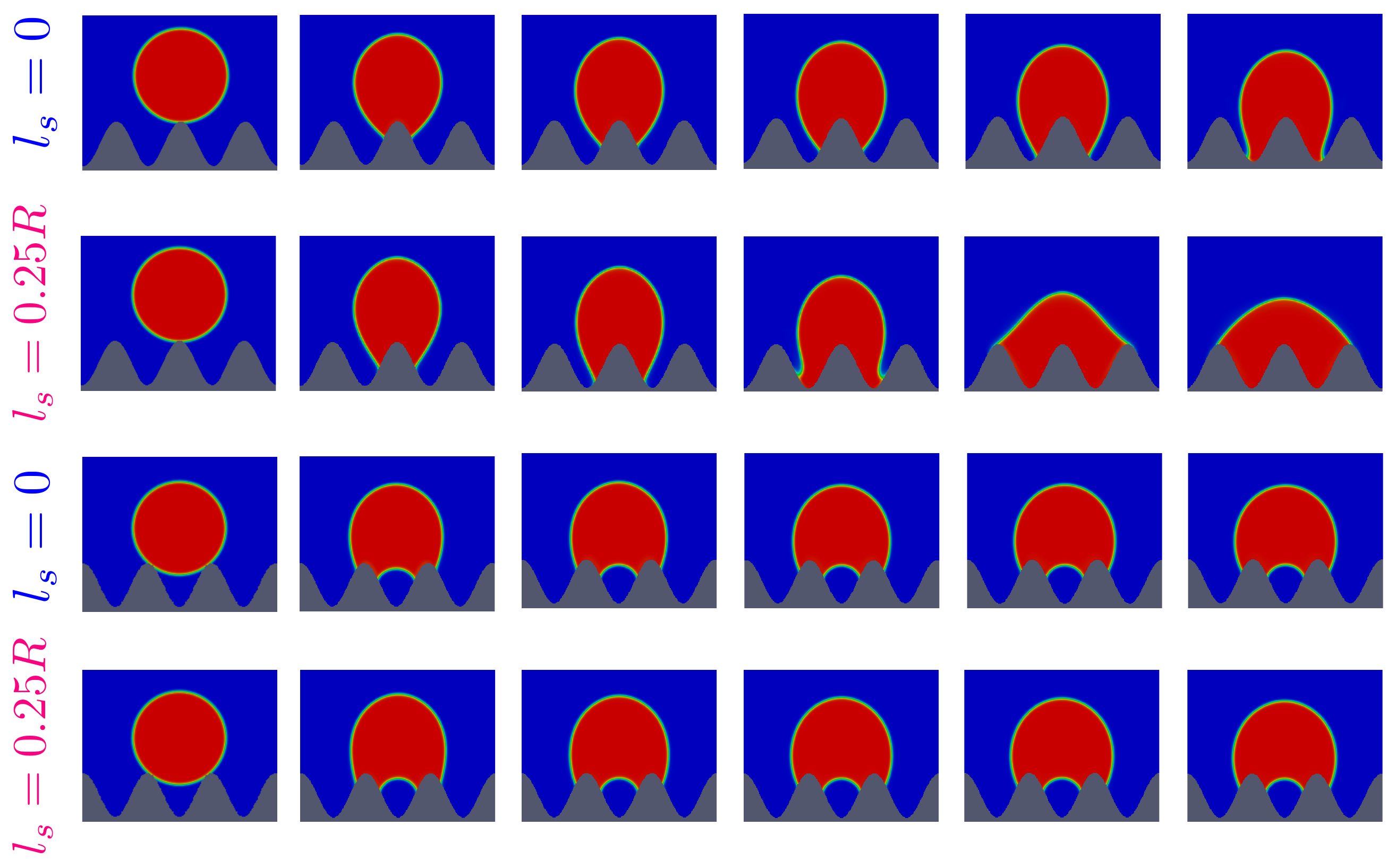}
\put(-272,-2){\small $t^*=0.$}
\put(-227,-2){\small $t^*\approx6.71$}
\put(-181,-2){\small $t^*\approx13.42$}
\put(-134,-2){\small $t^*\approx20.14$}
\put(-89, -2){\small $t^*\approx26.85$}
\put(-43, -2){\small $t^*\approx33.56$}
\put(-272,45){\small $t^*=0.$}
\put(-227,45){\small $t^*\approx6.71$}
\put(-181,45){\small $t^*\approx13.42$}
\put(-134,45){\small $t^*\approx20.14$}
\put(-89, 45){\small $t^*\approx26.85$}
\put(-43, 45){\small $t^*\approx33.56$}
\put(-272,92){\small $t^*=0.$}
\put(-227,92){\small $t^*\approx9.39$}
\put(-181,92){\small $t^*\approx18.79$}
\put(-134,92){\small $t^*\approx28.19$}
\put(-89, 92){\small $t^*\approx37.59$}
\put(-43, 92){\small $t^*\approx46.99$}
\put(-272,139){\small $t^*=0.$}
\put(-227,139){\small $t^*\approx9.39$}
\put(-181,139){\small $t^*\approx18.79$}
\put(-134,139){\small $t^*\approx28.19$}
\put(-89, 139){\small $t^*\approx37.59$}
\put(-43, 139){\small $t^*\approx46.99$}
\put(-310, 165){\large $a)$} 
\put(-310, 120){\large $b)$} 
\put(-310, 67){\large $c)$} 
\put(-310, 22){\large $d)$} 
\caption{Droplet spreading over two sinusoidal surfaces for two different initial droplet locations. The first and the third rows represent the results of spreading without any slip velocity whereas the second and the fourth rows show the results of droplet spreading over surfaces with slip length equal to  $l_s=0.25R_0$. }
\label{fig:SinusoidalSurfacesCont}
\end{figure}
\begin{figure} [H] 
\centering
\includegraphics[width = 0.48\textwidth]{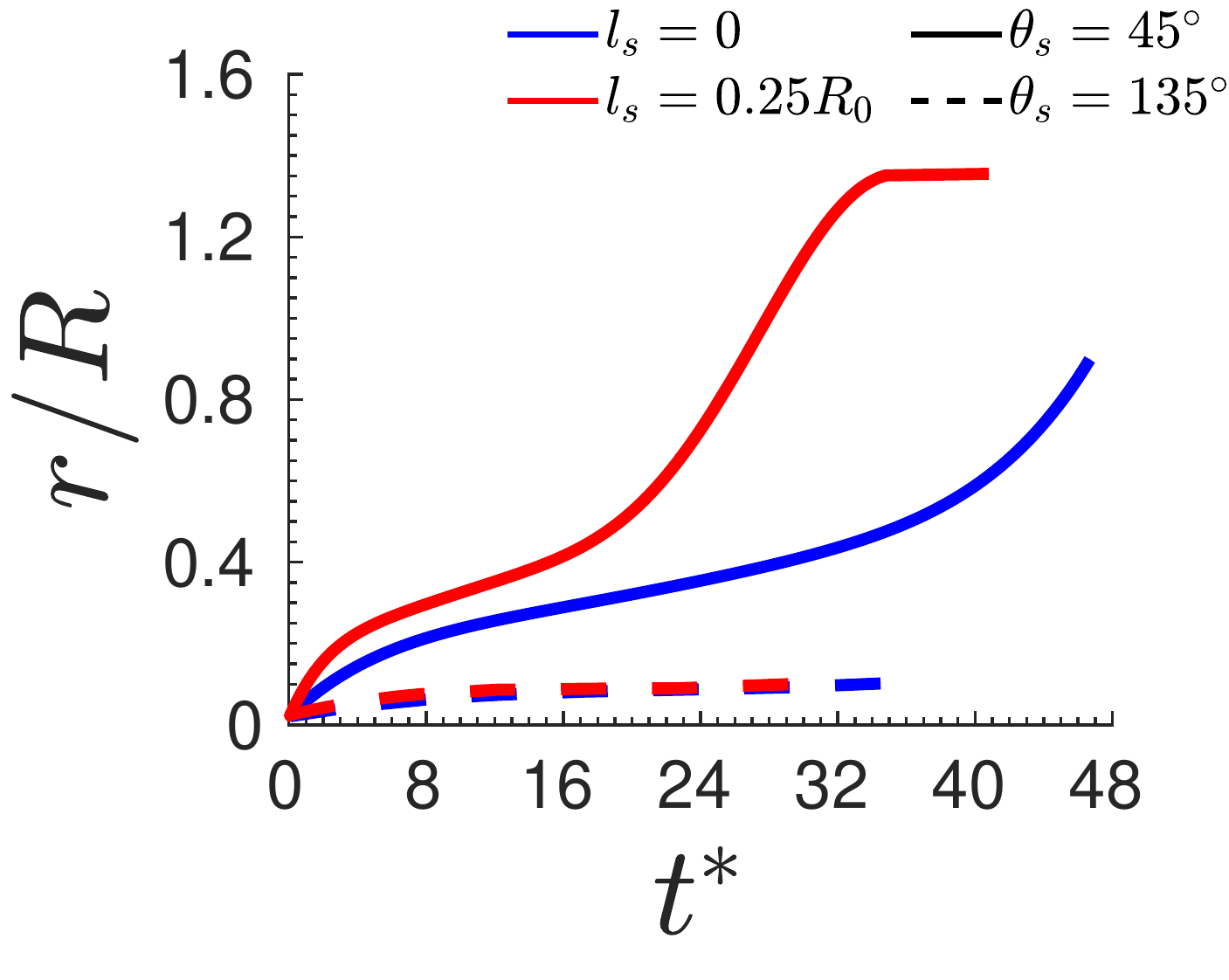}
\includegraphics[width = 0.51\textwidth]{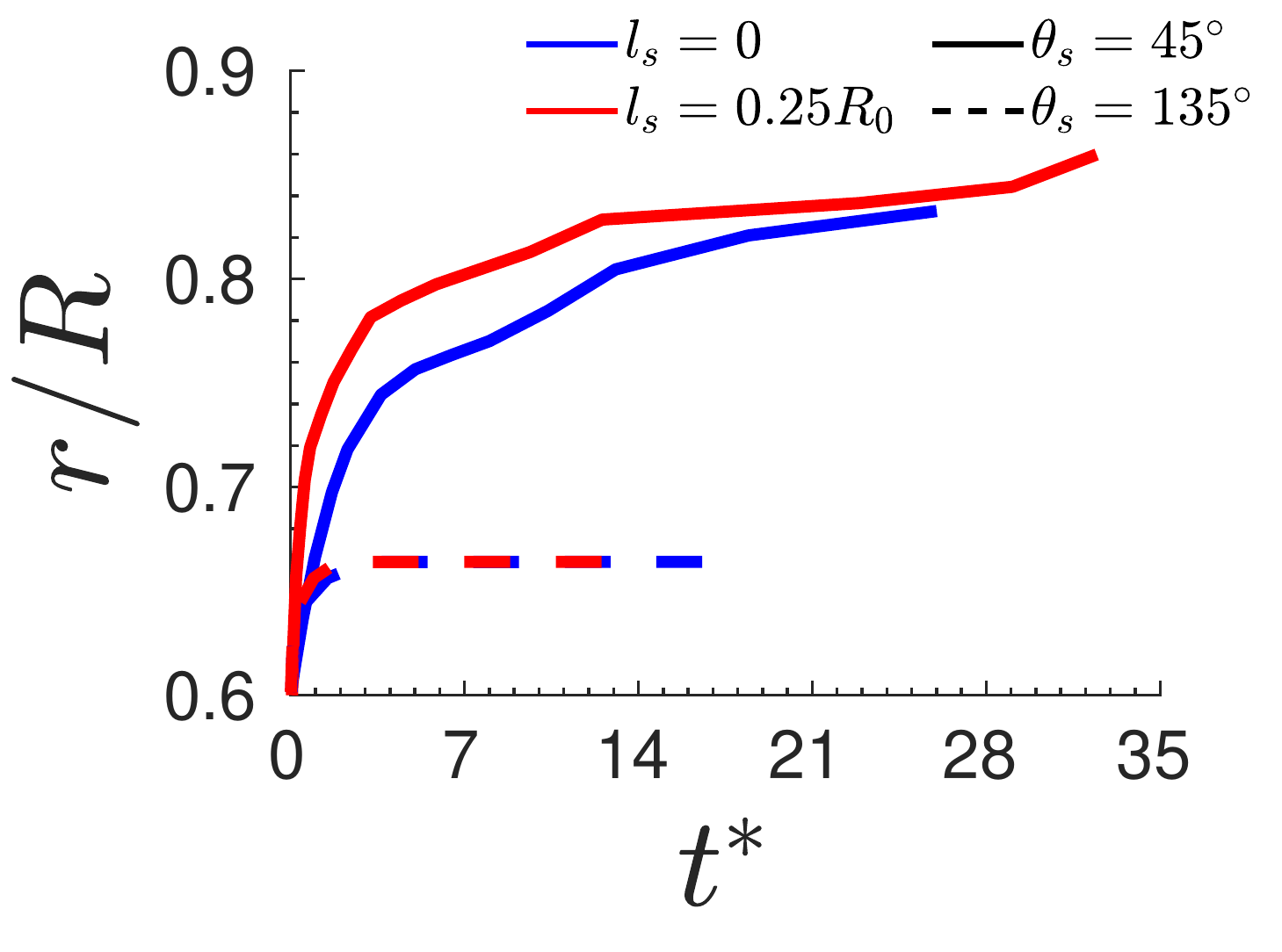}
\put(-340,131){\large a)}
\put(-160,131){\large b)}
\caption{Evolution of the droplet equivalent spreading radius for two different initial positions with the respect to the wall crests. The red and the blue colours indicate the cases with slip and no-slip velocity boundary conditions, whereas solid line denotes the results for a hydrophilic wall and the dashed-line those for the hydrophobic wall.}
\label{fig:SinusoidalSurfacesSpr}
\end{figure}

\begin{figure} [H] 
\centering
\includegraphics[width = 0.8\textwidth]{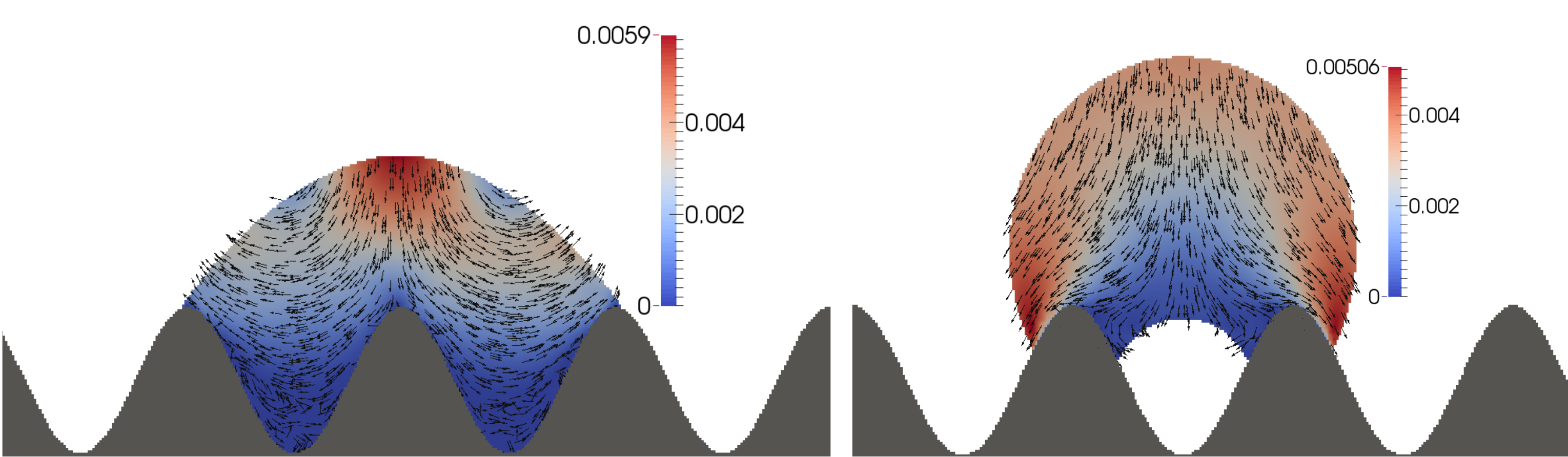}
\put(-270,60){\large a)}
\put(-120,60){\large b)}
\caption{Samples of  velocity contour inside the droplet for the cases with slip length, $l_s=0.25R_0$. Panels a and b represent the contour of velocity at $t^* \approx 46.99$ for the case $m=0$, and  $t^* \approx 33.56$ for $m=1$.} 
\label{fig:VelContour}
\end{figure}

Figure \ref{fig:SinusoidalSurfacesCont} shows the contour of the order parameter for two different wall geometries and two different slip lengths at six different time instances.

In this case, we define an equivalent spreading radius as the horizontal distance between two contact lines, reported in figure \ref{fig:SinusoidalSurfacesSpr}  normalised with the initial radius of the droplet (for $m=0$ in fig~\ref{fig:SinusoidalSurfacesSpr}a and $m=1$ in fig~\ref{fig:SinusoidalSurfacesSpr}b). 
The red and the blue colours indicate the cases with slip and no-slip velocity boundary conditions, respectively. The solid line denotes the results for the hydrophilic wall and the dashed-line those for the hydrophobic wall. 

Figures \ref{fig:SinusoidalSurfacesCont} and \ref{fig:SinusoidalSurfacesSpr} show that the wall geometry makes a significant difference in the droplet spreading. When the droplet is initially placed on a cavity (see fig~\ref{fig:SinusoidalSurfacesCont}c and fig~\ref{fig:SinusoidalSurfacesCont}d), the trapped gas inside the cavity cannot leave it; hence, to conserve the mass of the trapped gas, the droplet wets the surface by forming an arc shape. As a consequence the static contact angle is reached faster and the spreading is limited. Therefore, a static configuration is achieved faster without significant spreading. On the other hand, in the absence of any trapped gas (see fig~\ref{fig:SinusoidalSurfacesCont}a and fig~\ref{fig:SinusoidalSurfacesCont}b), the droplet fully wets the wall and spreads over the surface. The spreading  continues until the equilibrium contact angle is attained; however, given  the computational cost, we stopped the simulations at $t^*\approx 47$. In addition, we note that a slip velocity remarkably speeds up the spreading when the droplet fully wets the wall (compare fig~\ref{fig:SinusoidalSurfacesCont}a and fig~\ref{fig:SinusoidalSurfacesCont}b). Finally, as shown in figure \ref{fig:SinusoidalSurfacesSpr}, the spreading radius on a hydrophobic wall is much less than that on its hydrophilic counterpart. 

Figure \ref{fig:VelContour} depicts the velocity contours inside the droplet for the cases with slip length, $l_s=0.25R$. Panels a and b show the velocity contour at $t^*\approx 47$ for the case with the droplet initially placed on the wall crest, and  at $t^*\approx 33.5$  when the droplet is initially placed on the cavity. Note that the velocity vectors shown in figure \ref{fig:VelContour} are not scaled and only indicate the velocity direction. The magnitude of the velocity is displayed by the background colour.

\begin{figure} [H] 
\centering
\includegraphics[width = 0.49\textwidth]{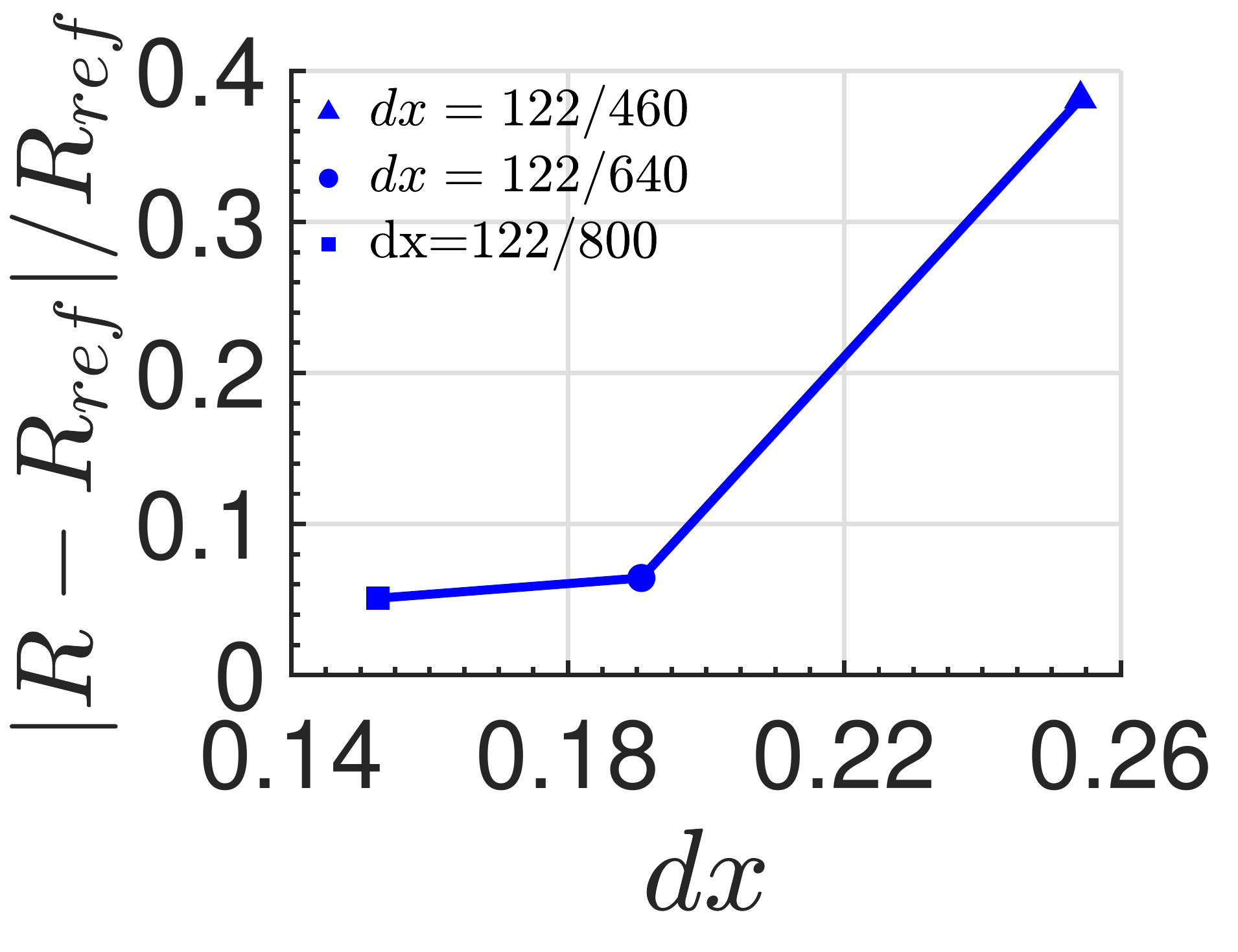}
\includegraphics[width = 0.49\textwidth]{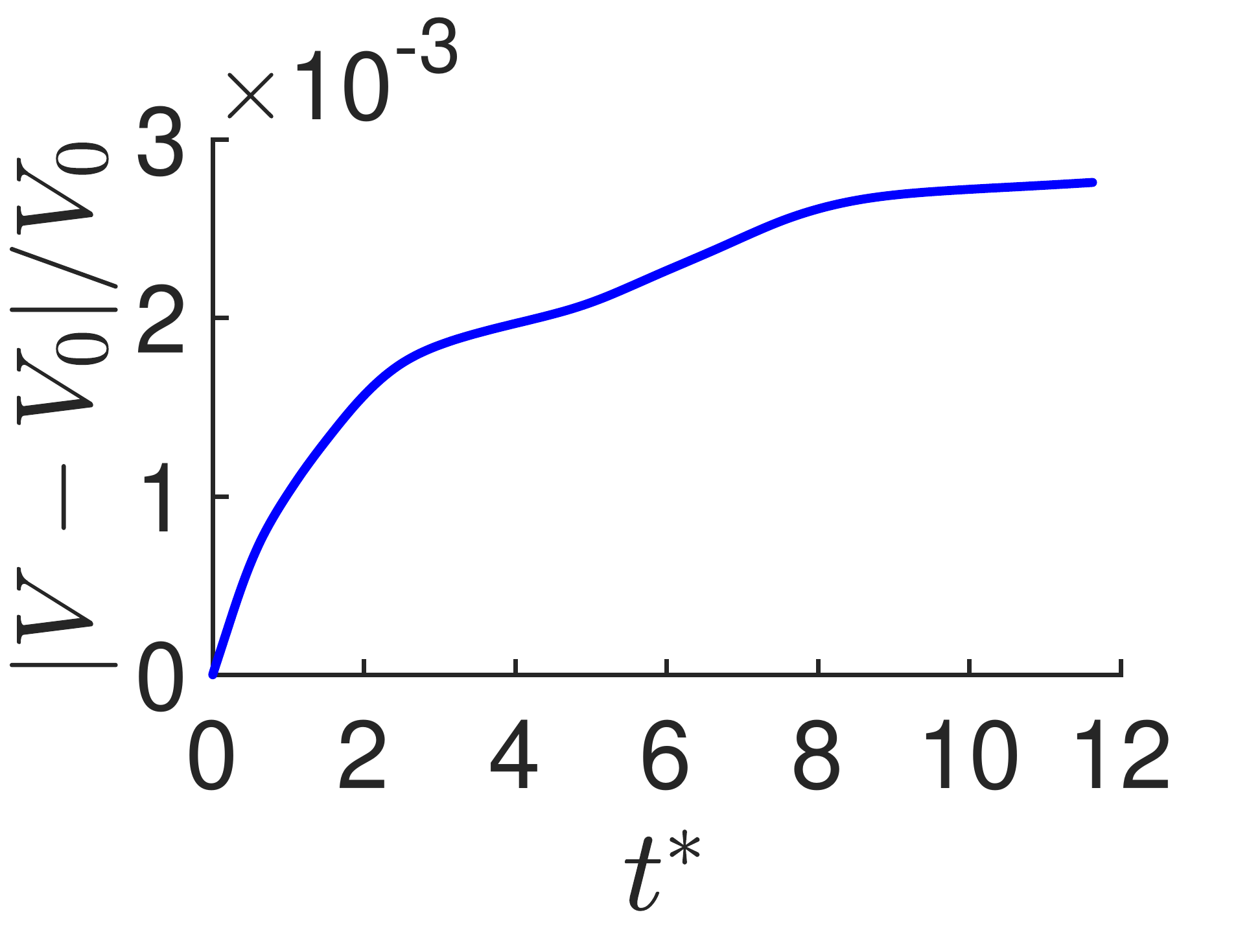}
\put(-210,120){\large a)}
\put(-30,120){\large b)}
\caption{Grid convergence for $l_s=0.25R$ and $m=0$ (left panel) and evolution of the droplet shrinkage for the converged grid size (right panel).}   
\label{fig:convergence}
\end{figure}
To examine the convergence of the algorithm,  we performed the same simulations as in figure \ref{fig:SinusoidalSurfacesCont}b for three additional grid sizes, namely, $2560 \times 1280$,  $1600 \times 800$, and $960 \times 480$. For all of the simulations, we calculate the evolution of the wetting radius. 
We consider the results of the finest grid size ($2560 \times 1280$) as the reference values ($R_{ref}$) and calculate the normalised error obtained with the coarser numerical grids, see 
gigure \ref{fig:convergence}a. Note that to reduce the computational costs, we performed the simulations up to $t^*=2$. According to figure \ref{fig:convergence}a, for the grid sizes equal to or smaller than $122/640$, the results of the simulations are almost independent of the grid size.  It is worth to mention that due to the combinations of the
 different parameters affecting the necessary grid size (geometry, velocity boundary condition, contact angle boundary condition, etc.),  a grid study should be performed for each specific problem under study.

Mass leakage is a well-known problem of any phase-field model \citep{HUANG2020109192}. Different numerical methods have been proposed for solving the Cahn-Hilliard equation and to reduce the droplet shrinkage.  In figure \ref{fig:convergence}b,  we report the evolution of the relative change in the volume of the droplet for a case with slip velocity boundary condition ($l_s=0.25R$) at the sinusoidal wall without any phase shift  ($m=0$) obtained with $1280\times640$ grid points.  Our result illustrates that up to $t^*=12$, the mass loss is less than $0.3 \%$ which shows that the proposed IBM algorithm does not introduce additional mass leakage.

\subsection{Three-dimensional droplet spreading }
As mentioned in section \ref{Sec:recipie}, the algorithm presented above can be extended to three-dimensional formulations. In this section, we present the results of a simulation performed to model the spreading of a three-dimensional initially spherical droplet over a three-dimensional surface. The simulation is performed using the semi-implicit algorithm with a time step equal to $10^{-4}$.\\
A droplet of radius $R$ is initially placed at $[X_1,X_2,X_3]=[4R,4R,3.185R]$. The solid wall is generated as a portion of a sphere with radius $R_w= 9R$ and centre located at $[X_1,X_2,X_3]= [4R,4R,-27R/4]$.  The simulation domain is  $ [0,8R]\times[0,8R]\times[0,8R]$ (discretised with $640\times640\times640$ grid points). The non-dimensional parameters of these simulations are $Re=4$, $ca=1$, $Pe=10^4$, $Cn= 0.022$, $\theta_{eq}=45^\circ$, $l_s=0$, and $\mu_f=0$. \\

\begin{figure} [H] 
\centering
\includegraphics[width = 0.8\textwidth]{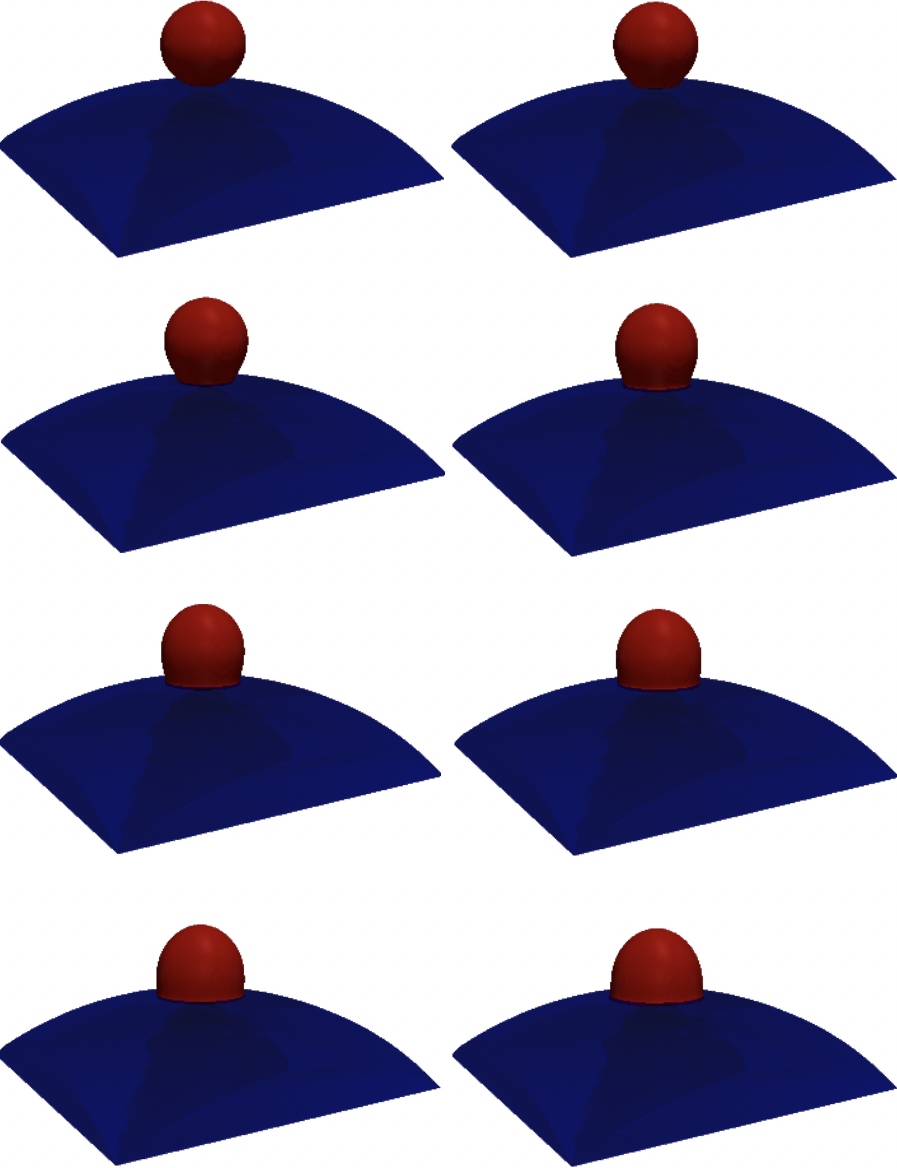}
\put(-90, -9){\small $t^*\approx 82.56$}
\put(-90, 85){\small $t^*\approx 58.97$}
\put(-90, 179){\small $t^*\approx 35.38$}
\put(-90, 271){\small $t^*\approx 11.79$}
\put(-230, -9){\small $t^*\approx 70.76$}
\put(-230, 85){\small $t^*\approx 47.17$}
\put(-230, 179){\small $t^*\approx 23.58$}
\put(-230, 271){\small $t^*=0$}
\caption{Spreading of a three-dimensiolan droplet over a three-dimensional wall.}
\label{fig:3D}
\end{figure}

\begin{figure} [H] 
\centering
\includegraphics[width = 0.65\textwidth]{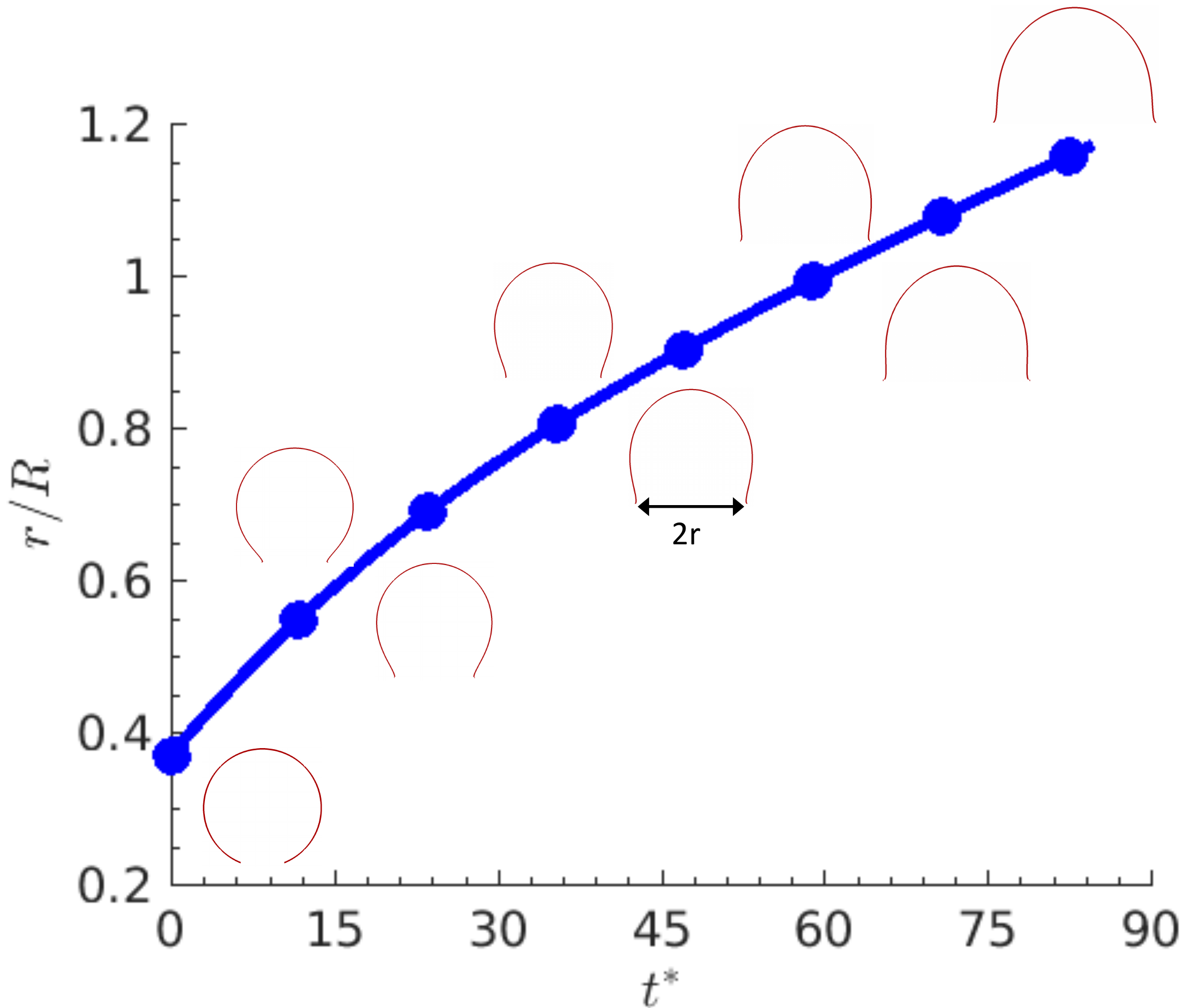}

\caption{Time evolution of the equivalent spreading radius of a three-dimensional droplet over a three-dimensional wall. The filled circles and the red contours represent the equivalent spreading radii and the interface contours at eight time instants of figure  \ref{fig:3D}}
\label{fig:3Dspreading}
\end{figure}

Figure \ref{fig:3D} shows the evolution of the droplet spreading over the surface at eight different time instants. The time evolution of the equivalent normalised wetting radius is presented in figure \ref{fig:3Dspreading}. Filled circles and red contours show the wetting radii and the two-dimensional cross-sections of the interface at the same times as the images figure \ref{fig:3D}. 

The results of this simulation prove the capability of the algorithm to model three-dimensional droplet spreading over any arbitrary stationary wall.  Moreover, since all the variables required for the IBM treatment (coordinates of ghost points, normal vectors at the wall, coordinates of averaging points, and averaging weights are stored in permanent arrays, the IBM module of the code does not add significant computational cost to the base PFM solver. 

\section{Conclusion} \label{sec:conclusion}
We have presented a fully Eulerian hybrid immersed-boundary phase-field model for simulating contact line dynamics on any arbitrary fixed solid wall. The algorithm consists of two independent modules, namely, a phase field and an immersed boundary module. The Navier-Stokes and Cahn-Hilliard equations are solved on a cartesian numerical mesh yet imposing  boundary conditions on any complex wall geometries
 using a volume-penalisation and ghost-cell immersed boundary method. The proposed algorithm is capable modelling both static and dynamic-contact angle boundary conditions with possibly slip velocity at the wall. The proposed algorithm has the following novel properties:
\begin{itemize}
\item The fully Eulerian approach facilitates an efficient implementation and, in particular, parallelisation and accelerated architectures.
\item It consists of an initialisation step during which all the  auxiliary quantities necessary for the immersed boundary treatment of the complex wall are calculated and stored (coordinates of ghost points, normal vectors at the wall, coordinates of averaging points, and averaging weights).  Hence, the IBM module does not add significant computational cost to the base solver for the system formed by the Navier-Stokes and Cahn-Hilliard equations.
\item It suits both two- and three-dimensional simulations, without additional complexities in three-dimensions.
\item Due to the modular feature of the algorithm, the phase field formulation, particularly the free energy of the system, can be modified with no effect on the solution algorithm, Thus the same strategy presented here can be employed to model different near-wall physical phenomena (for instance solidification over a complex wall geometry).
\end{itemize}

The numerical tests reported in this manuscript validate the algorithm against different results from the literature on wetting and two-fluid systems.Hence, the proposed algorithm can be see as an efficient and powerful method to study multiphase flows near solid boundaries using free-energy formulations, in particular contact line dynamics over complex geometries. However, the IBM module presented here, with the accurate and efficient calculations of normal and tangential vectors, interpolation and extrapolation through an immersed boundary, can be used to impose arbitrary mixed boundary conditions for flow problems over complex geometries, not only for single phase flows but also multiphase flow simulations using  other Eulerian approaches to track an interface, e.g.\ volume-of-fluid and level-set methods. 
Examples of simulations where the approach proposed here could be of help are large-eddy and RANS simulations where wall models are necessary \citep{Roman,Amitabh} and heat-transfer problems where boundary conditions involve both the temperature and concentration field as well as their gradients \citep{LUPO2019118563}.

\appendix \label{Appendix}
\section{ Semi-implicit algorithm}
The semi-implicit algorithm follows the same steps as the fully explicit one but with different numerical procedure for  the Cahn-Hilliard and Navier-Stokes equations. In this appendix the semi-implicit algorithms are presented as these turn out to be more stable and  to allow for a longer time step, so they are preferable for an efficient implementation when dealing with larger problems.

\subsection{Cahn-Hilliard equation}
We follow the idea of  \cite{YUE}  and \cite{DONG2012} to decompose the Cahn-Hilliard equation, see equation \ref{Cahn-Hilliard} in the main text, into two Helmholtz equations. First we solve equation \ref{Helmholtz1} for the auxiliary variable $\Gamma$. 
\begin{linenomath}\begin{equation} \label{Helmholtz1}
 \nabla^2 \Gamma -(\alpha_s + \frac{S}{\epsilon^2})\Gamma = \frac{1}{\lambda M}\left(\frac{\check C}{\Delta t} \\
-\frac{\partial}{\partial x_i}(\tilde{u_i}^{n+1} \tilde{C}^{n+1}) \right)+ \nabla^2 \left[\frac{1}{\epsilon^2}  \psi(\tilde{C}^{n+1})-\frac{s}{\epsilon^2}\tilde{C}^{n+1} \right]
\end{equation}\end{linenomath}
where $\lambda= \frac{3}{2\sqrt{2}}\sigma \epsilon$ is the mixing energy density. In equation \ref{Helmholtz1}, for any arbitrary variable $\chi$ we have a first estimation at time $n+1$ denoted  by $\tilde{\chi}^{n+1}$.  To achieve a second order accuracy in time, we estimate the time derivative at time $n+1$ as $(3/2\chi^{n+1}-\check{\chi})/dt$, where     $\tilde{\chi}^{n+1}$ and $\check{\chi}$ are defined as follows \citep{DONG2012}:
\begin{linenomath}\begin{equation}
\check{\chi} = 2\chi^n-\frac{1}{2}\chi^{n-1}, \tilde{\chi}^{n+1} =  2\chi^n-\chi^{n-1}.
\end{equation}\end{linenomath}
$S$ is the stabilisation parameter and is chosen such that $S\geq \epsilon^2 \sqrt{\frac{6}{\lambda M \Delta t}}$. 

The coefficient $\alpha_s$ of the second Helmhotz equation is also computed following \cite{DONG2012}, i.e.\
\begin{linenomath}\begin{equation}
\alpha_s = -\frac{S}{2\epsilon^2}\left( 1+\sqrt{1-\frac{6\epsilon^4}{\lambda M \Delta t S^2}}\right).
\end{equation}
Next we solve equation a second Helmhotz problem of the form 
\begin{equation} \label{Helmholtz2}
\begin{gathered}
 \nabla^2 C^{n+1} + \alpha_s C^{n+1} = \Gamma,
 \end{gathered}
\end{equation}\end{linenomath}
and update the value of the order parameter at time $n+1$.

Here, we solve both equations \ref{Helmholtz1} and \ref{Helmholtz2} by taking Fourier transforms. Note that the boundary conditions for the order parameter and the 
velocity at the wall are imposed through the IBM algorithm. Therefore, in the wall-normal direction $X_2$ direction, we simply consider Neumann boundary conditions for both $\Gamma$ and $C$ together with the no-slip boundary condition for the velocity, whereas periodic boundary conditions are considered in the flow  direction, $X_1$, for all the variables.

\subsection{Navier-Stokes equations}
For the semi-implicit approach, we again use a fractional step method. First we calculate the first and the second prediction velocities, $u^*$ and $u^{**}$. Following the idea of  \cite{DONG2012} and \cite{DODD2014416}, we solve the following Helmholtz equation for  $u^*$,
\begin{linenomath}\begin{equation} \label{predictionSemi}
\begin{gathered}
\frac{u^*_i-u^n_i}{\Delta t} = -\left( \frac{3}{2} \frac{\partial}{\partial x_j}(u^n_i u^{n}_j)-\frac{1}{2} \frac{\partial}{\partial x_j}(u^{n-1}_i u^{n-1}_j) \right)+
\left(  \frac{3}{2} \frac{\phi^{n+1}}{\rho^{n+1}}\frac{\partial C^{n+1} }{\partial x_i}- \frac{1}{2} \frac{\phi^{n+1}}{\rho^{n+1}}\frac{\partial C^{n+1} }{\partial x_i}    \right)+\\
\frac{1}{\rho^{n+1}}\left[ \frac{\partial}{\partial x_j}\left(\mu^{n+1}(\frac{\partial u^n_i}{\partial x_j}+\frac{\partial u^n_j}{\partial x_i})  \right) \right] +\\
\frac{1}{2} \left(\nu_0 \nabla^2 u^*-\nu_0 \nabla^2 u^n \right)+\\
\left( \frac{3}{2}  f^n_{b_i}-\frac{1}{2}  f^{n-1}_{b_i} \right)+\frac{\rho_1-\rho_2}{2}M\frac{\partial \phi^{n+1}}{\partial x_j}\frac{\partial u_i^n}{\partial x_j},\\
u^{**}_i = u^{*}_i \alpha
 \end{gathered}
\end{equation}\end{linenomath}
 where $\nu_0 = \frac{1}{2} \left( \frac{\mu_1}{\rho_1}+\frac{\mu_2}{\rho_2} \right)$. Note that the last term in equation \ref{predictionSemi} is added to consistently conserve the mass flux at the interface  \cite[see][for a detailed discussion]{HUANG2020109192}.
Finally, we update $u_{n+1}$ and $p^{n+1}$ similarly to what done with the explicit algorithm,
\begin{equation} \label{Correction}
\frac{u_i^{n+1}-u^{**}}{\Delta t} = (\frac{1}{\rho_0}-\frac{1}{\rho^{n+1}})\frac{\partial P^{n+1}}{\partial x_i}.
\end{equation}

\section*{Acknowledgements}
The research was financially supported by the Swedish Research Council, via the multidisciplinary research environment INTERFACE (VR  2016-06119 ``Hybrid multiscale modelling of transport phenomena for energy efficient processes''). The computation resources were provided by SNIC (Swedish National Infrastructure for Computing) and by the National Infrastructure for High Performance Computing and Data Storage in Norway (project no. NN9561K). MER was supported by the JSPS KAKENHI Grant Number JP20K22402.
 
\bibliography{bibfile}

\end{document}